\documentclass{aastex}
\usepackage{spr-astr-addons}
\usepackage{url}

\RequirePackage{color}

\begin{document}

\title{VARIATION OF THE TULLY-FISHER RELATION AS A FUNCTION OF THE MAGNITUDE INTERVAL OF A SAMPLE OF GALAXIES.}

\author{A. Ruelas-Mayorga\altaffilmark{1}, L. J. S\'anchez\altaffilmark{1}, M. Trujillo-Lara\altaffilmark{1}, A. Nigoche-Netro\altaffilmark{2}, J. Echevarr\'{\i}a\altaffilmark{1}, A. M. Garc\'{\i}a\altaffilmark{1}, J. Ram\'{\i}rez-V\'elez\altaffilmark{1}}

\altaffiltext{1}{Instituto de
Astronom\'{\i}a, Universidad Nacional Aut\'onoma de M\'exico,
M\'exico D.F., M\'exico.}
\altaffiltext{2}{Instituto de Astronom{\'\i}a y Meteorolog{\'\i}a, Universidad
     de Guadalajara, Guadalajara, Jal. 44130, M\'exico.}

\shortauthor{Ruelas-Mayorga, S\'anchez, Trujillo-Lara, Nigoche-Netro, Echevarr\'{\i}a and Garc\'{\i}a}
\shorttitle{Tully-Fisher Relation}

\abstract{In this paper we carry out a preliminary study of the dependence of the Tully-Fisher Relation (TFR) with
the width and intensity level of the absolute magnitude interval of a limited sample of 2411 galaxies taken from Mathewson \& Ford (1996).
The galaxies in this sample do not differ significantly in morphological type, and are distributed over an $\sim11$-magnitude
interval ($-24.4 < I < -13.0$). We take as directives the papers by Nigoche-Netro et al.
(2008, 2009, 2010) in which they study the dependence of the Kormendy (KR), the Fundamental Plane (FPR) and the
Faber-Jackson Relations (FJR) with the magnitude interval within which the observed galaxies used to derive these
relations are contained. We were able to characterise the behaviour of the TFR coefficients $(\alpha, \beta)$ with respect to the width of the magnitude interval as well as with
the brightness of the galaxies within this magnitude interval. We concluded that the TFR for this specific sample of galaxies depends
on observational biases caused by arbitrary magnitude cuts, which in turn depend on the width and intensity of the chosen brightness levels.}


\maketitle

\section{Introduction}

\label{sec:intro}

There are studies published in the literature (Nigoche-Netro et al.
2008, 2009, 2010) in which the authors carry out studies of the dependence of
the parameter values of structural relations of elliptical galaxies on the
size and brightness of the magnitude interval in which the observed galaxies
are contained. The structural relations which are studied in these papers are:
the Kormendy Relation (KR), the Fundamental Plane Relation (FPR) and the
Faber-Jackson Relation (FJR). The authors point out that it may not be possible
to reach conclusions on the physical properties of groups of
galaxies when comparing the values of the slopes of the structural relations
if these values were obtained with galaxy samples contained in magnitude
intervals of different width or in different magnitude intervals.
If the magnitude
intervals are narrow, the differences encountered are negligible, whereas, if
the interval is wide, the geometric form of the galaxy distribution on a plane
formed by the structural relations parameters is dominated by the magnitude
cuts induced by the observations, and this could mask the differences produced
by the galaxies intrinsic properties.

A study similar to the ones described above for spiral galaxies has not been
performed. This paper presents a preliminary study of whether
these effects are also present in the TFR of a limited sample of galaxies which do not differ
significantly in morphological type (Sb-Sc).
In this paper we investigate
whether arbitrary observational magnitude cuts to spiral galaxy samples
produce biases on the TFR parameter values. If this is so, these parameter
values may not be used in calculations of the distance of spiral galaxies
using the TFR. It is important to stress at this point that the aim of this paper
is not so much obtaining the true values of the coefficients of the TFR, but rather
showing that these values may change depending on the way the data used in calculating
the parameters are collected.

In 1977 R. Brent Tully and J. Richard Fisher (Tully \& Fisher, 1977) published a
paper in which they established a relation between the total luminosity and the rotational velocity ($L=\beta V_{rot}^{\alpha}$) for several samples of spiral
galaxies. From this relation, and the
maximum value of the rotational velocity, it is possible to estimate the
absolute magnitude of the entire galaxy, and hence, by comparison with the
apparent magnitude, calculate its distance.

Therefore the TFR is a very important tool to map the large scale
structure of the universe, and the Hubble flow.

For nearby galaxies the TFR requires $HI$ observations (recessional
velocity $\leq10,000$ $km/s$), whereas for larger recessional velocities we
use $H\alpha$ observations, although when the recessional velocities are larger
that $60,000$ $km/s$ $(z=0.2)$, $H\alpha$  appears with wavelength
longer than $7875\mathring{A}$ where it may be confused with $OH$ sky emission lines. For
larger distances the use of other emission lines, such as: [OII] (3727 \AA ) and
[OIII] (5007 \AA ) is normal (Vogt et al., 1996). These authors point  to the fact that
the TFR may be used to study the structure and evolution of spiral galaxies
with large systemic velocities, due to the fact that their luminosities (found
using the TFR) do not vary much $(\Delta B_{M}\leq0.6)$ with respect to
those found by spectroscopic means for more distant galaxies (Bamford et al, 2006).

The observations of spiral galaxies used for establishing the parameters $(\alpha, \beta)$ of
the TFR must be corrected for a variety of effects, among which some of the
most important ones are the correction for inclination, the correction for
internal absorption and also the correction for dust extinction within our own galaxy.

If we consider, as it is usually done, that spiral galaxies may be represented
by oblate spheroids, the inclination may be calculated from the disc
projection using the following equation:

\bigskip

\bigskip
\begin{center}
$\cos^{2}i=\frac{\left(  \frac{b}{a}\right)  ^{2}-\alpha^{2}}{1-\alpha^{2}}$
\end{center}

\bigskip

where $i$ represent the angle of inclination, $\frac{b}{a}$ the minor to major
axis ratio of the best fitted ellipse, and $\alpha$ is the intrinsic axial
ratio for an edge-on system. There are some problems, however, because some
galaxies do not have perfectly circular isophotes when viewed face-on, so we
must restrict our galaxy sample to inclinations of at least $35^{o}-45^{o}$ in
order to minimise the uncertainties on the deprojected values of rotational
velocity.

To determine the internal extinction in galaxies has always been very
difficult. However, at present, the use of CCD and IR detectors allows very
precise multiwavelength surface photometry for galaxies, which in turn,
permits the statistical determination of reddening. The extinction corrections
are expressed in terms of the axial ratio \ (Tully et al, 1998, Masters,
Giovanelly \& Hanes, 2003):

\bigskip

\begin{center}
$A_{\lambda}^{i}=\gamma_{\lambda}\log(\frac{a}{b})$
\end{center}

\bigskip

where

\bigskip

$A_{\lambda}^{i}$ represents the extinction at wavelength $\lambda$ as a
function of the inclination angle $i$, and $(\frac{a}{b})$ \ the observed
axial ratio.

This correction depends on the dust content of galaxies as variations of the
coefficient $\gamma_{\lambda}$ as a function of the rotational velocity
corrected for inclination effects $\left(  W_{R}^{i}\right)  $.

Recently Shao et al. (2007) have made an extensive study of a sample of more than 60,000 galaxies from the second Data Release of the SDSS.
They find that the Luminosity Function (LF) of spiral galaxies appears to be consistent with a simple obscuration model; for which the optical depth
$\tau$ results to be proportional to the cosine of the inclination angle $(i)$. This cosine function is multiplied by a coefficient $(\gamma)$ which
turns out to be independent of the luminosity of the galaxy in question in any one of the bands studied, giving a power law of wavelength as an extinction curve
$(\tau \sim \lambda^ {-n})$ where the exponent takes a value of $n=0.96 \pm 0.04$. The characteristic magnitudes $(M^*)$ of the LF are made dimmer by
$0.5$ mags in the $z$ band and $1.2$ mags in the $u$ band. Since the Mathewson \& Ford (1996) sample that we use for this paper has already been corrected
for a variety of effects, including inclination corrections, we do not attempt any further corrections in order to preserve the consistency of the
data we are using.

Apparent magnitudes also require corrections for dust effects in our own
galaxy. Using the $HI$ and dust maps (IRAS 100 $\mu m$) we know that the amount
of $HI$ and that of dust are well correlated. Extinction due to this components
has been calibrated with observations of distant stars in different colours.
As a result the galactic extinction for any galaxy may be estimated given its
position on these maps  (Schlegel, Finkbeiner \& Davis, 1998). If the galactic
latitude is lower than $25^{o}-30^{o}$, corrections for galactic extinction
become larger as well as the associated uncertainties (Schlafly \& Finkbeiner,
2011). This is why the samples of galaxies to which the TFR is applied are
usually restricted to those with $l\geq25^{o}$. However, there is a recent paper
(Said, Kraan-Korteweg \& Jarrett, 2015) in which the authors have found a correction in the
Near Infrared (NIR) that allows the inclusion of galaxies located at lower galactic
latitudes, in the so-called Zone of Avoidance (ZoA).

\bigskip

The TFR has been absolutely calibrated by the use of Cepheids in external
galaxies, observed with the Hubble Space Telescope (HST), whose distances
extend out to $\sim20$ $Mpc$ (see Freedman, 1990 and Pierce \& Tully, 1992.).
One such study is the \textit{Hubble Key Project of Extragalactic Distances},
which obtained distances, using Cepheids,  to galaxies that contain type Ia
Supernovae.

\bigskip

In Vogt et al. (1996), the TFR has been used to calculate the distance to many
spiral galaxies with systemic velocities smaller than $8,000$ $km/sec$, the
results of this paper show a very good linear relation between the distance in
$Mpc$ and the receding velocity of the galaxies in $km/sec$, showing quite clearly the
Hubble flow and producing a value for the Hubble constant of 80 $kms^{-1}$ $Mpc^{-1}$.

\bigskip

Since its discovery in 1977, the TFR has been a very powerful tool, used to calculate the distance to galaxies whose
position very far from our vantage point makes it difficult to obtain their distances using other more conventional methods, as well as to
provide insight into the formation and evolution of galaxies.

As mentioned above, the TFR is a relation between the brightness of a spiral galaxy and its rotational velocity, this brightness is measured in different passbands. Due to this fact, there are TFRs in the radio for HI and CO observations as well as for optical and infrared bands. The general form of the TFR is the same for all passbands but the values of its coefficients and scatter may vary.

Although it appears that the TFR is valid for spiral galaxies at different redshifts, a change of the values of its parameters has been detected and is referred to as evolution of the TFR, Ziegler et al. (2002) have studied 60 late-type galaxies in the redshift interval $0.1-1$. They find that the more distant sample presents a flatter TFR than that for the local galaxies, being the values they find $-5.77 \pm 0.45$ and $-7.92 \pm 0.18$, thus finding evidence of evolution of the TFR at $3\sigma$ levels. Weiner, et al. (2006) measure the evolution of the TFR in the optical and infrared using kinematic measurements of a large sample of galaxies from the Team Keck Redshift Survey on the GOODS-N field. They detect evolution in both the slope and the intercept of the TFR, this results suggest differential luminosity evolution. With the aid of the multi-integral field spectrograph GIRAFFE at the VLT, Puech, et al. (2008) have derived the K-band TFR at $z \sim 0.6$ for a sample of 65 galaxies. They conclude that both the slope and the scatter of the TFR do not appear to evolve with redshift. Fern\'andez, et al. (2009) study the $B$, $R$ and $I$ TFR at $z=1.3$. Their results are not conclusive in suggesting evolution of the TFR, since the possible luminosity evolution is
contained within the scatter of the relation. Their study, however, shows a clear tendency for all bands studied favouring a luminosity evolution where galaxies were brighter in the past for the same rotational velocity. Fern\'andez, et al. (2010) study the evolution of the TFR in the $B$, $V$, $R$, $I$, and $K_s$ bands. They detect a clear evolution of the TFRs in the sense that galaxies were brighter in the past at the same rotational velocity. The luminosity change is more noticeable for shorter wavelengths.

Russell (2004) has found that the TFR pre\-sent a dependence on galactic morphology. Sc I and Seyfert galaxies are more luminous at a given rotational velocity than galaxies of other morphological types. Not taking into account this difference may lead to the distances to Sc I galaxies to be systematically underestimated, whereas distances to Sb/Sc III galaxies are overestimated. He concludes that using type-dependent TFR improves significantly the determination of distances to galaxies. Masters, et al. (2008) have investigated the dependence of the NIR TFR with morphological type and have shown that for the $J$, $H$, and $K$ bands the TFR is shallower for earlier-type spirals which also have a brighter TFR zero-point than the later-type spirals. Shen, at al. (2009) show that the TFR of spiral galaxies is very morphological-type dependent, where earlier-type spirals have systematically lower luminosities at a fixed maximum rotational velocity. This difference is more pronounced at shorter wavelengths.

As we said above, the TFR was discovered in 1977 (Tully \& Fisher, 1977) and since this date, it has
been widely utilised to study peculiar velocities and cosmography  (e.g. Han, 1992; Mould et al., 1993; da Costa et al., 1996; Theureau et al., 2007;
 Courtois \& Tully, 2012; Courtois et al., 2013; Tully et al., 2014). The rotational velocity, which is distance independent, may be measured using either
 optical rotation curves or $HI$ profiles line-width, and the absolute magnitude of these galaxies may be obtained through photometric measurements. Once this is done in a consistent way, a truly reliable TF survey is obtained. This is exactly what was done by Mathewson \& Ford, (1996) for a particular sample of more than 2,000 galaxies perfectly suited for the TFR. This sample is the one we use in this preliminary study (see Section \ref{sec:sample}). If the results
 obtained in this paper look promising we plan to extend this preliminary study to a much larger statistically significant study using suitable spiral galaxies from the Sloan Digital Sky Survey (SDSS).

\bigskip

In this paper we shall report the results of a preliminary study of the dependence of the
value of the coefficients of the TFR with the size and width of the magnitude
interval within which the observed galaxies are contained. In \S2 we present a sample of 2411 galaxies which we use for this study, in
\S3 and the Appendix we perform a series of analysis for the data which may be conceptualised as different observational arrangements for the galaxies used in the calculation of the TFR coefficients, \S4 presents  a non-parametric statistical analysis with which we prove that the variation of the values of the coefficients is
not due to statistical fluctuations and therefore, may be ascribed to the way the galaxies have been grouped together, \S5 and the Appendix present an analysis of the values of the TFR coefficients with apparent magnitude, and \S6 presents our conclusions.

\section{THE SAMPLE OF GALAXIES}
\label{sec:sample}

We have used a sample of 2411 galaxies of the 2447 studied by Mathewson \& Ford (1996), the 36 galaxies difference is due to the fact that
those galaxies lacked one or several of the parameters necessary for the calculations and fits that we perform in this paper. $I$-band luminosities, rotational velocity and redshifts of 1092 of these spiral galaxies were measured by the authors using the 1-m and 2.3-m telescopes at the Siding Spring Observatory, they used CCD photometry and $H\alpha$ spectroscopy, and the results were reported in their 1996 paper. They combined these data with similar data for 1355 galaxies previously published in Mathewson, Ford \& Buchhorn (1992). These galaxies were selected from the ESO-Uppsala Catalogue of types Sb-Sc galaxies, with diameters between $1^{\arcmin}.0$ and $1^{\arcmin}.6$, and velocities between $4,000$ and $14,000$ $kms^{-1}$, inclinations greater than $40^o$ and galactic latitude above $b=+ 11^o$ or below $b= -11^o$. For full details on the samples, the data reduction and the table of data, see the papers cited in this paragraph.

The data in Mathewson \& Ford (1996) represent an ideal sample of galaxies already used by the authors in the TFR, besides all the data published in their table are already corrected for  internal and external extinction, inclination, K-correction, etc, which make their use really straight forward for our purposes.

From their catalogue we used the apparent $I$ magnitude, the redshift and the maximum $H\alpha$ rotational velocity. We calculated the absolute $I$ magnitude by deriving the distance to each galaxy using the non-relativistic relation $D=cz/H$ where $D$ stands for the distance in $Mpc$, $c$ is the speed of light in $km/s$, $z$ is the redshift and $H$ is the Hubble constant. We use $H=70$ $kms^{-1}Mpc^{-1}$ (Shen et al., 2009).

\begin{figure}
\begin{center}
\includegraphics[width=8.0cm,height=8.0cm]{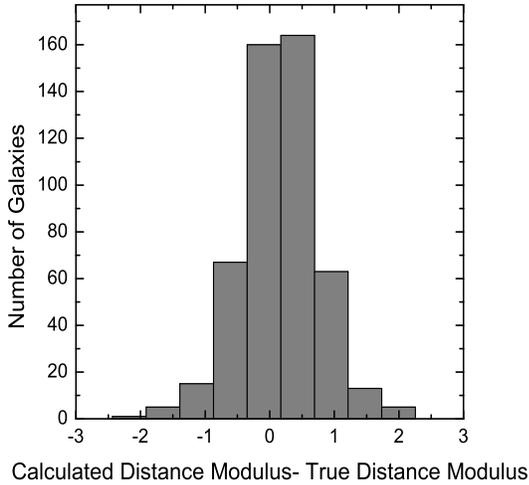}
\end{center}
\caption{Distribution of the difference of the values of the distance moduli (DM) for a subsample of our sample of galaxies. The true DM was obtain using the true distance from the HyperLeda catalogue, and the calculated DM was obtained from the simple formula $D=cz/H$. $454$ galaxies out of a total of $493$ are contained within the 4 central bars ($-0.87 \ mags$  to $+0.69 \ mags$).} \label{fig:HISTOGRAM}
\end{figure}

Since the galaxies used in this sample are not very distant $(D<250 \ Mpc)$, the distance derived as indicated in the previous paragraph $D=cz/H$ is not very reliable. To gauge how much difference there would be between the real $I$ absolute magnitude calculated with the distance derived with this simple formula, and the real value corresponding to the true distance, we searched the HyperLeda  \footnote{http://leda.univ-lyon1.fr/} catalogue for galaxies in our sample for which their true distance is determined using Cepheid variables. We found $493$ such galaxies in our sample. We made a histogram (see Figure~\ref{fig:HISTOGRAM}) of the difference between the distance modulus obtained using the formula $D=cz/H$ and the real distance modulus. The histogram was symmetrical around $0 \ mags$, and $92.1 \%$ of the galaxies $(454)$ were contained from $-0.87 \ mags$  to $+0.69 \ mags$, giving a mean deviation of $\sim \pm 0.8 \  mags$. If we assume that the behaviour of the entire sample is similar to that of this subsample, we may conclude that the formula $D=cz/H$ would produce a maximum deviation of the value of the $I$ absolute magnitude calculated for the galaxies in the sample of $\sim \pm 0.8$ mags, and that the deviations would be distributed uniformly around the zero value, so statistically they would not bias the determination of the values of the TFR parameters one way or the other.

For the TFR we used the expression  published in Shen et al. (2009) which is as follows:

\begin{center}
$M=\alpha\log\left(  \frac{V_{\max}}{200\  km/s}\right)  +\beta$.
\end{center}

\begin{figure}
\begin{center}
\includegraphics[width=8.0cm,height=8.0cm]{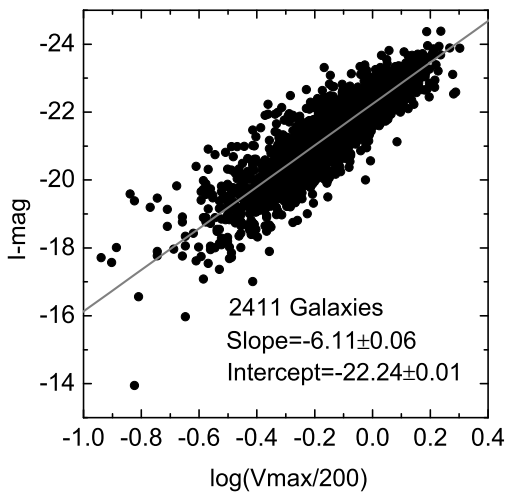}
\end{center}
\caption{Least squares fit to the 2411 galaxies used in this paper. The values of the slope and the intercept are shown on the figure as well as their errors.} \label{fig:TODASFIT}
\end{figure}

In Figure~\ref{fig:TODASFIT} we see the plot of these 2411 galaxies on the I (absolute I magnitude) versus $log(V_{max}/200)$ plane, as well as the resulting fitted line and the fit parameters.

\bigskip

\section{ANALYSIS OF DIFFERENT CASES}
\label{sec:cases}

It is well known that the value of the parameters of the TFR vary according to the galactic type of the galaxies in question
(Tully \& Fisher (1977), Giraud (1986), Russell (2004), Masters (2008), Shen et al. (2009), Said et al. (2015)). Mathewson \& Ford (1996) state that all the galaxies in their paper are of types Sb-Sc. So, for this preliminary study
we expect the variation of the values of the $\alpha$, and $\beta$ coefficients due to the different galactic types to be negligible, however, at this point we cannot exclude that part of the variations
that we encounter in this paper may be due to the difference in galactic type.  It is important to mention that Russell (2004) finds that not using a type-dependent TFR will result in underestimated distances for Sb/Sc III galaxies and overestimated distances for Sc I galaxies, this effect will probably cancel itself out in the determination of the TFR coefficients leading, at most, to a weak dependence on morphological type for the sample used in this paper.

\begin{figure}
\begin{center}
\includegraphics[width=8.0cm,height=8.0cm]{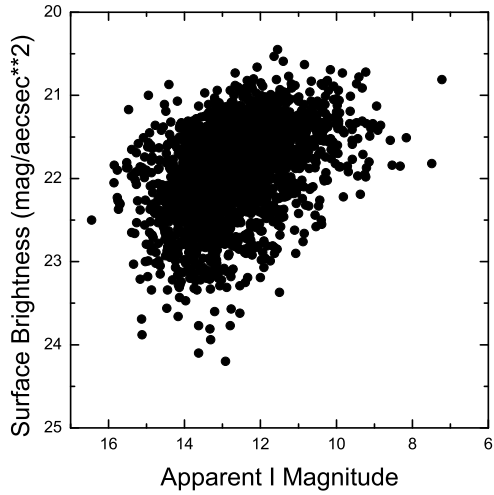}
\end{center}
\caption{Surface brightness versus Apparent I magnitude.}\label{fig:SURF_BRIGHT_APP}
\end{figure}

\begin{figure}
\begin{center}
\includegraphics[width=8.0cm,height=8.0cm]{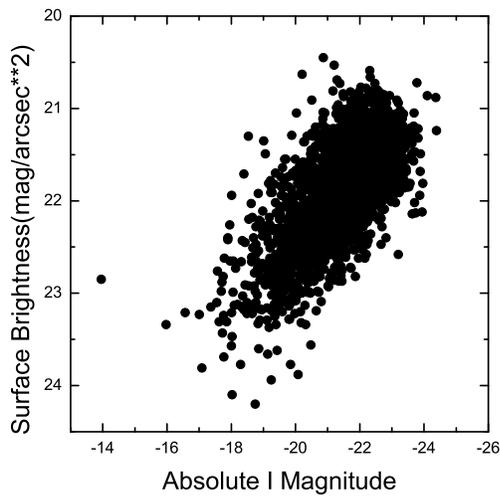}
\end{center}
\caption{Surface brightness versus Absolute I magnitude.}\label{fig:SURF_BRIGHT_ABS}
\end{figure}

It is well known that ``there is weak evidence for a surface brightness dependency" of the TFR coefficients (Tully, 2007). We present in
Figures~\ref{fig:SURF_BRIGHT_APP} and \ref{fig:SURF_BRIGHT_ABS} plots of the surface brightness of the galaxies in our sample versus
apparent and absolute I magnitude. We see that in the magnitude interval of interest for this paper, the surface brightness of our sample of galaxies
varies approximately between $\sim 23.5$ and $\sim 20.5$ $mag/arcsec^2$, being the typical width of the distribution, at a fixed magnitude value,
 $\sim 2$ $mag$ for the apparent magnitude case and $\sim 1.5$ $mag$ for the absolute magnitude case. We consider the size of our sample too
 small for conducting a serious study of the possible dependence of the values of the coefficients of the TFR with surface brightness. We shall
 pursue such study in the future with a larger sample of galaxies. At present we cannot exclude the possibility that part of the changes we
 observe in the value of the TFR coefficients may be due to a weak dependence with the surface brightness.

In this paper, we study the entire range of data magnitudes in a number of cases which we report in what follows. The points that represent the galaxies in each case have been fitted using a least squares bisector fit, that means that the final results -slope and intercept- correspond to those for the line that bisects the lines obtained performing a normal least squares fit using the rotational velocity as independent variable first, and then using the absolute magnitude as independent variable. This procedure follows closely that used in the Nigoche et al. (2008), (2009), and (2010) papers.

The values of the TFR coefficients $(\alpha, \ \beta)$ were studied in five different cases which are summarised in Table \ref{tab:cases}.

\begin{table*}
\caption{Cases Studied} \label{tab:cases} \centering
\setlength{\tabcolsep}{0.5\tabcolsep}
\begin{tabular}{|c|c|c|c|c|}
\hline
Case          &    Type                                     &     $\Delta M$                           &  Starting Magnitude         &     Finishing Magnitude \\
\hline
              &                                             &                                          &                             &                         \\
     1        &    Fixed Magnitude Intervals                &          1                               &        -25                  &           -13            \\
     2        &    Increasing Width Magnitude Intervals     &          1                               &        -25                  &           -13             \\
     3        &    Increasing Width Magnitude Intervals     &          1                               &        -13                  &           -25             \\
     4        &    Increasing Width Magnitude Intervals     &          1                               &        -25.5                &           -13.5             \\
     5        &    Fixed Magnitude Intervals                &          2                               &        -25                  &           -13            \\
\hline
\end{tabular}
\end{table*}

\begin{figure}
\begin{center}
\includegraphics[width=8.0cm,height=8.0cm]{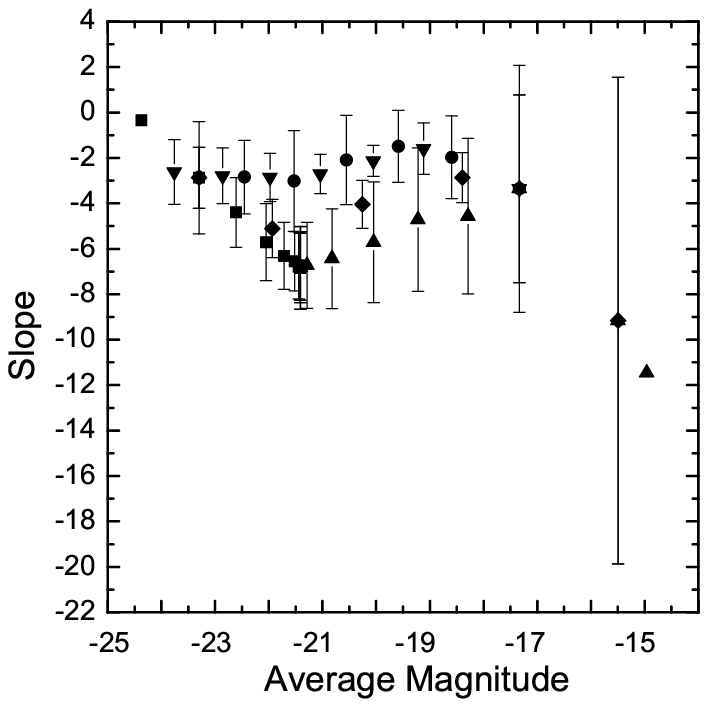}
\end{center}
\caption{Average values of the slope for the 5 cases discussed in section \ref{sec:cases}. Case 1: dots, Case 2: squares, Case 3: Triangles, Case 4: Inverted triangles and Case 5: Diamonds.} \label{fig:ALLSLOPE1}
\end{figure}

\begin{figure}
\begin{center}
\includegraphics[width=8.0cm,height=8.0cm]{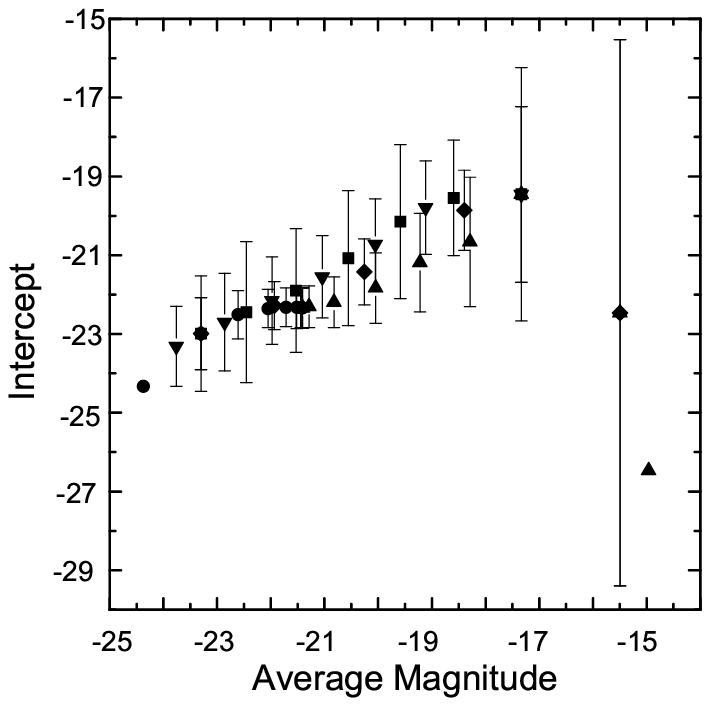}
\end{center}
\caption{Average values of the intercept for the 5 cases discussed in section \ref{sec:cases}. Case 1: dots, Case 2: squares, Case 3: Triangles, Case 4: Inverted triangles and Case 5: Diamonds.} \label{fig:ALLINTERCEPT1}
\end{figure}

In Figures \ref{fig:ALLSLOPE1} and \ref{fig:ALLINTERCEPT1}, we present, on the same plane, the values of the slope as well as those of the intercept for the five cases discussed above. These Figures allow us to appreciate visually the variation in the values of these parameters for the different cases.

In the Appendix we present tables with the values of coefficients $\alpha$ and $\beta$ for each one of the cases discussed above. From these tables it is clear that in some of the intervals the number of galaxies is rather small, therefore the coefficient determinations are not statistically significant within these intervals (One interval for Case 1, one for Case 2, five for Case 3, two for Case 4 and one for Case 5). We, however, leave in this paper these determinations for consistency with the analysis method established in Nigoche-Netro et al. (2008, 2009 \& 2010).

\section{Hypothesis tests for $\alpha$ and $\beta$}
\label{sec:hypothesis}

In this paper we present a statistical test of the variation of the values of the TFR coefficients in order to establish whether their value variation is real or whether it corresponds only to statistical fluctuations. We shall use a non-parametric test known as the $Run$ $Test$ (see Bendat \& J. S., Piersol, A. G., 1966).

The essence of this method is as follows: we consider $N$ observations of a random variable $x(k)$, where each observation is classified in one of two classes, which we shall identify with the $+$ and $-$ signs. A simple example of this may be the outcome of a series of coin tosses, where one side of the coin is identified with the $+$ sign, while the other with the $-$ sign. A $Run$ is a series of identical outcomes which is followed or preceded by a different observation or by no observations at all.

It is proven that in a sequence of $N$ independent observations of the same random variable, the distribution of the variable $r(k)$ has a mean value and a variance determined by the following equations:

\bigskip

\begin{center}
$\mu_{r}   =\frac{2N_{1}N_{2}}{N}+1$
\end{center}

\vskip0.5cm

\begin{center}
$sigma_{r}^{2}   =\frac{2N_{1}N_{2}(2N_{1}N_{2}-N)}{N^{2}(N-1)}$
\end{center}

\bigskip

where $N_1$ is the number of $+$ and $N_2$ is the number of $-$, and $N=N_1 + N_2$.

One important condition of all non-pa\-ra\-me\-tric tests is the fact that no distribution function is assumed for the variable under study. The Run-test applied to cases where the two outcomes are independent, and have the same probability of appearing (0.5), generates a new variable (number of runs) which turns out to be distributed in a Gaussian manner. Therefore, if we set a limiting probability, say $99 \% $, we are able to determine an interval for the number of runs that have a probability of appearing of $99 \% $ or higher from the Gaussian distribution for the number of runs. If the number of runs we measure is within this interval the conclusion is that the probability of the two outcomes is equal to 0.5 and independent from one outcome to the next, however, if the measured number of runs does not fall within this interval, then we must conclude that the probability of the two outcomes is not equal or not independent from one outcome to the next, that is, there is an underlying tendency determined, in this case, with a $99 \% $ certainty level. In the case of the $\alpha$ and $\beta$ parameters we applied the Run test in the following manner. From the set of different values determined for, say $\alpha$, we calculated its median. Those values larger than the median were identified with the $+$ sign, while those smaller than the median with the $-$ sign. We then obtained the number of runs for this set of measurements. In all cases we were able to determine that there is an underlying tendency with the levels of significance given in Table \ref{tab:hypothesis}.

As explained above, we make use of this test to establish whether or not there is an underlying tendency in the results we obtain for the values of $\alpha$ and $\beta$. The procedure followed assumes, as a {\it null hypothesis}, that there is {\bf NO} underlying tendency in the variations of the values of $\alpha$ and $\beta$, application of this method permits us to reject the {\it null hypothesis} with a rather large level of confidence. Table \ref{tab:hypothesis} shows the percentages at which we can reject the {\it null hypothesis} for each one of the cases discussed in Section \ref{sec:cases}.

\begin{table}
\caption{Levels of rejection of the {\it null hypothesis} } \label{tab:hypothesis} \centering
\setlength{\tabcolsep}{0.5\tabcolsep}
\begin{tabular}{|c|c|c|}
\hline
Case          &    \% of rejection                          &     \% of rejection \\
              &     for Slope $(\alpha)$                    &           for Intercept $(\beta)$          \\
\hline
              &                                             &                                             \\
     1        &               98                            &                  99                         \\
     2        &               99                            &                  99                         \\
     3        &               99                            &                  99                         \\
     4        &               98                            &                  98                         \\
     5        &               69                            &                  93                         \\
\hline
\end{tabular}
\end{table}

Therefore, since we can reject the {\it null hypothesis} --there is no underlying tendency in the changes of the values of the coefficients of the TFR-- with such high degrees of confidence, then we may conclude that the variation of the values of these coefficients is {\bf NOT} due to random fluctuations, but rather to the particular observational conditions under which the data were collected.

At this point we must be reminded that this is a preliminary study performed on a small, for current standards, sample of galaxies. We, therefore, consider that having levels of certainty in the rejection of the null hypothesis of $\ge 2.3 \sigma$ represent a high degree of confidence considering the size of our sample and the preliminary nature of our study.

\section{Variations of $\alpha$ and $\beta$ with apparent magnitude}
\label{sec:magnitudecases}

In the previous sections we have studied the variation of the $\alpha$ and $\beta$ coefficients for different cases in which we have collected the galaxies
in absolute magnitude intervals. In doing so, we are making sure that the galaxies chosen in each interval have similar characteristics since we may assume
that they have similar values for their masses.

In what follows we shall undertake a similar study but collecting the galaxies in apparent magnitude intervals, this to the end of studying how the values
of the $\alpha$ and $\beta$ coefficients change as we take deeper observations of the sky. In a way we may say that this is an observational approach undertaken with telescopes which become each time more powerful, allowing us to detect fainter and fainter galaxies.

\begin{figure}
\centering
\includegraphics[width=8.0cm,height=8.0cm]{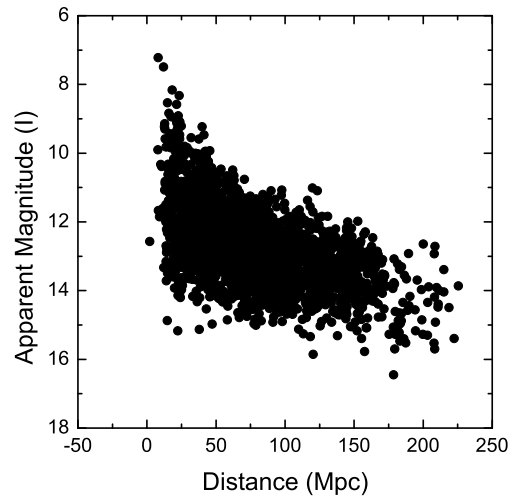}\caption{Distribution of the apparent magnitude of the galaxies in our sample as a function of distance.}\label{fig:appmag_vs_dist}
\end{figure}

In Figure \ref{fig:appmag_vs_dist} we present the distribution of the apparent magnitude of the galaxies in our sample as a function of distance.

From this figure it is clear that those galaxies that appear bright in our observations are also located nearby, but as we include apparently fainter galaxies
we see that the distance interval covered by them gets larger, until at an apparent magnitude of $i \sim 12$ the galaxies occupy a distance interval which is much larger. This interval goes from $0$ $Mpc$ to $ \sim 200$ $Mpc$.

\begin{figure}
\centering
\includegraphics[width=8.0cm,height=8.0cm]{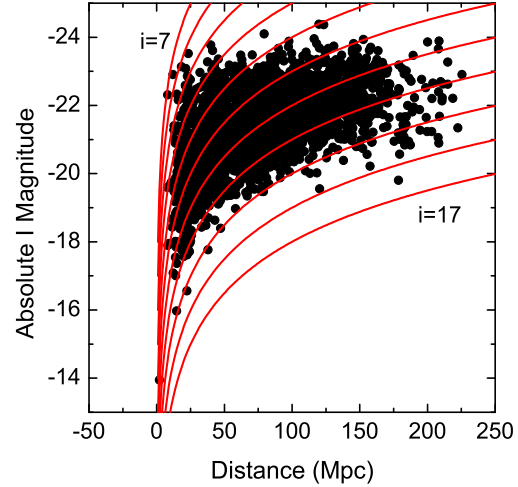}\caption{Absolute I Magnitude versus Distance (Mpc). The curved lines represent the loci of constant apparent magnitude starting at $i=+17$ (bottom curve) and ending at $i=+7$ (top curve) in steps of $-1$ $mag$.}\label{fig:MALMQUIST}
\end{figure}

Since we are collecting the galaxies in apparent magnitude intervals, we shall be treading the classical realm of the Malmquist bias. To establish whether
our sample of galaxies suffers from this bias we shall present in Figure~\ref{fig:MALMQUIST} a plot of Absolute I Magnitude versus Distance in Mpc. On this diagram we also plot curves of constant apparent $i$ magnitude, which go from $i=+17$ (bottom curve) to $i=+7$ (top curve) in steps of $-1$ magnitude.

``The infallible signature of bias in any sample is an apparent brightening of the individual absolute magnitude with increasing distance in a flux-limited
sample." (Sandage, 2000). It is quite clear from Figure~\ref{fig:MALMQUIST} that for the cases in which we take intervals of fixed-width, as well as for those
in which we take increasing width intervals the brightening of the sample as we move towards larger distances is very clear. Therefore, we must conclude that
our sample suffers from Malmquist bias which must be corrected for. If we assume that the galaxies are distributed uniformly in space, and that their magnitudes are distributed around a corresponding mean value with the same dispersion $(\sigma)$, then the average magnitude calculated must be dimmed by an amount equal to $1.382\sigma^2$ (see Sandage, 2000). This will show on the graphs for parameter values (Figures~\ref{fig:between10and11} to \ref{fig:between14and15}, and Figures~\ref{fig:HASTA10} to \ref{fig:HASTA17}) as a shifting of the points to fainter (larger) values of magnitude, without changing the value of the slope $(\alpha)$, but affecting the value of the intercept $(\beta)$ which will experience a change of its value given by $+1.382\sigma^2$. This correction was not performed in our results simply because we lack a reliable value for $\sigma$, however since the purpose of this paper is not so much to calculate the values of the TFR parameter but instead to show that their values vary depending on the observational cuts imposed on the data, not applying the Malmquist bias correction is irrelevant for the purposes pursued herein.

In what follows we shall discuss the results of calculating the TFR parameters $(\alpha, \beta)$ in apparent magnitude intervals as indicated in Table \ref{tab:intervals}, $n$ represent the number of galaxies in that apparent magnitude interval. The graphs corresponding to these cases are presented in the Appendix.

\begin{figure}
\centering
\includegraphics[width=8.0cm,height=8.0cm]{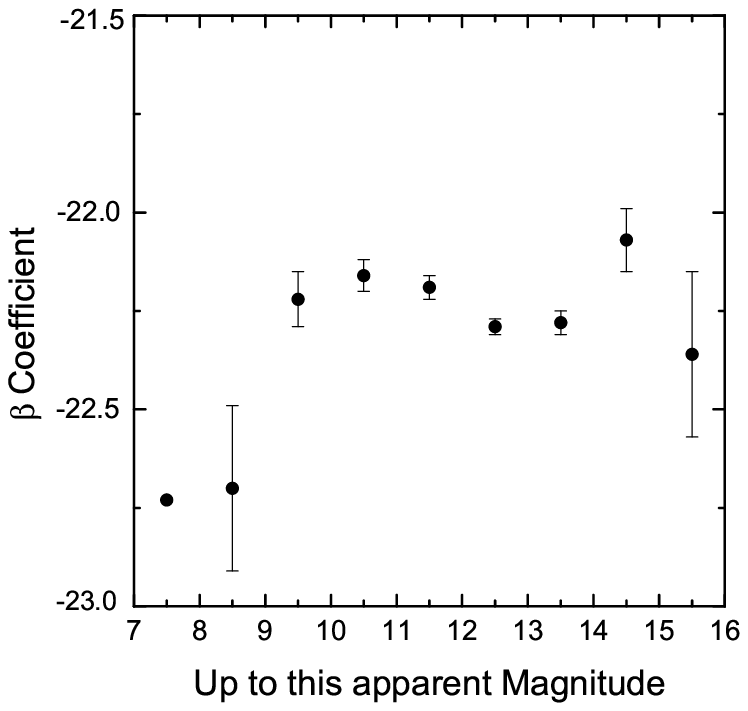}\caption{Variations of the value of the $\beta$
coefficient for groups of galaxies contained in intervals 1 apparent magnitude wide.} \label{fig:beta1mag1}
\end{figure}

In Figure \ref{fig:alpha1mag1} we see that the value
of $\alpha$ presents variations which are larger than the size of the error bars, however, its value stays contained within a
band that extends from $-5$ to $-7$. Figure \ref{fig:beta1mag1} shows the values of the $\beta$ coefficient. This coefficient also
presents variations larger than the size of the typical error bars, but its value remains contained within a band extending from
$-22.0$ to $-22.5$.

\begin{figure}
\centering
\includegraphics[width=8.0cm,height=8.0cm]{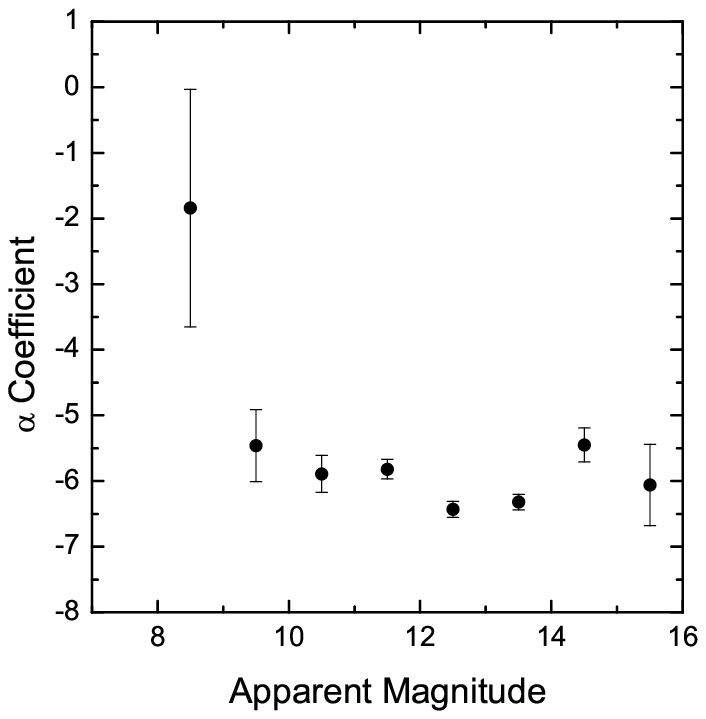}\caption{Variations of the value of the $\alpha$
coefficient for groups of galaxies contained in intervals 1 apparent magnitude wide.} \label{fig:alpha1mag1}
\end{figure}

\bigskip

\begin{table}
\caption{Galaxies in Intervals of Apparent Magnitude }\label{tab:intervals}\centering
\setlength{\tabcolsep}{0.5\tabcolsep}
\begin{tabular}
[c]{|c|c|c|c|c|c|}\hline
Apparent Magnitude & $\alpha$ & $\Delta\alpha$ & $\beta$ & $\Delta\beta$ &   $n$\\
Interval           &          &                &         &               &      \\
\hline
&     &        &     &       & \\
7-8   & -21.02 & --- & -22.73 & ---  & 2\\
8-9   & -1.84 & 1.81 & -22.70 & 0.21 & 7\\
9-10  & -5.46 & 0.55 & -22.22 & 0.07 & 45\\
10-11 & -5.89 & 0.28 & -22.16 & 0.04 & 144\\
11-12 & -5.82 & 0.15 & -22.19 & 0.03 & 412\\
      &       &      &       &      &    \\
12-13 & -6.43 & 0.12 & -22.29 & 0.02 & 748\\
13-14 & -6.32 & 0.12 & -22.28 & 0.03 & 765\\
14-15 & -5.45 & 0.26 & -22.07 & 0.08 & 256\\
15-16 & -6.06 & 0.62 & -22.36 & 0.231 & 31\\\hline
\end{tabular}
\end{table}

At this time, we shall present a similar analysis in which galaxies will be accumulating, starting with those galaxies contained within a
bright apparent magnitude interval, until we reach the galaxies in the faintest apparent magnitude interval. We begin with the brightest galaxies in our sample up to an apparent magnitude of $+8$, continuing with all those galaxies with an apparent magnitude fainter than $+9$, until we reach
the total sample at an apparent magnitude of $+17$.

The values of the $\alpha$ and $\beta$ coefficients found in this analysis are shown in Table \ref{tab:valuesaccumulated}, as well as their errors
and the number of galaxies used in the calculations.

\begin{table}
\caption{Accumulated Galaxies in Intervals of Apparent Magnitude }\label{tab:valuesaccumulated}\centering
\setlength{\tabcolsep}{0.5\tabcolsep}
\begin{tabular}
[c]{|c|c|c|c|c|c|}\hline
Up to & $\alpha$ & $\Delta\alpha$ & $\beta$ & $\Delta\beta$ & $n$\\\hline
&  &  &  &  & \\
8 & -21.02 & --- & -22.73 & --- & 2\\
9 & -1.99 & 1.58 & -22.68 & 0.16 & 9\\
10 & -5.22 & 0.55 & -22.28 & 0.07 & 54\\
11 & -5.80 & 0.25 & -22.19 & 0.03 & 198\\
12 & -5.82 & 0.13 & -22.19 & 0.02 & 610\\
   &       &      &       &      &    \\
13 & -6.12 & 0.09 & -22.24 & 0.02 & 1358\\
14 & -6.18 & 0.07 & -22.25 & 0.01 & 2123\\
15 & -6.13 & 0.07 & -22.24 & 0.01 & 2379\\
16 & -6.11 & 0.06 & -22.24 & 0.01 & 2410\\
17 & -6.11 & 0.06 & -22.24 & 0.01 & 2411\\\hline
\end{tabular}
\end{table}

\begin{figure}
\centering
\includegraphics[width=8.0cm,height=8.0cm]{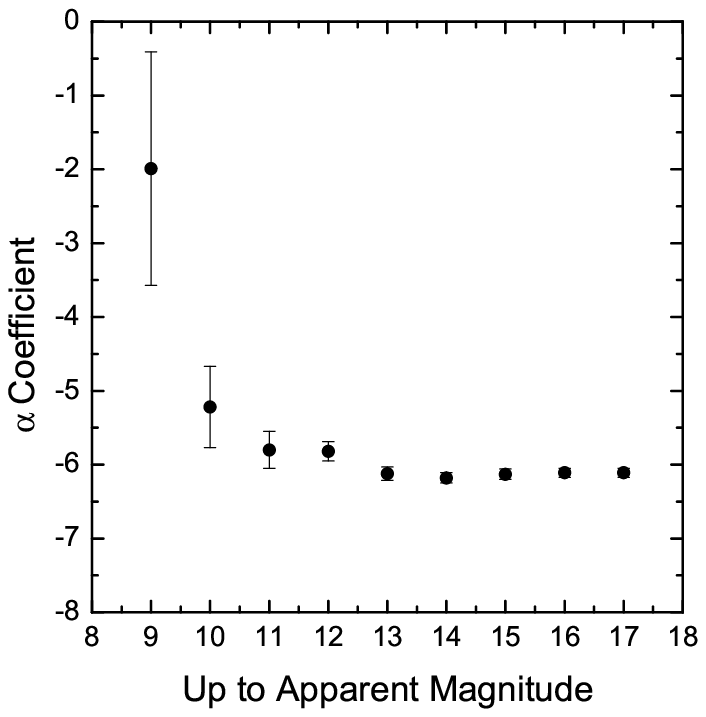}\caption{Variation of the $\alpha$ coefficient as a function of apparent magnitude.} \label{fig:alpha2mag1}
\end{figure}

\begin{figure}
\centering
\includegraphics[width=8.0cm,height=8.0cm]{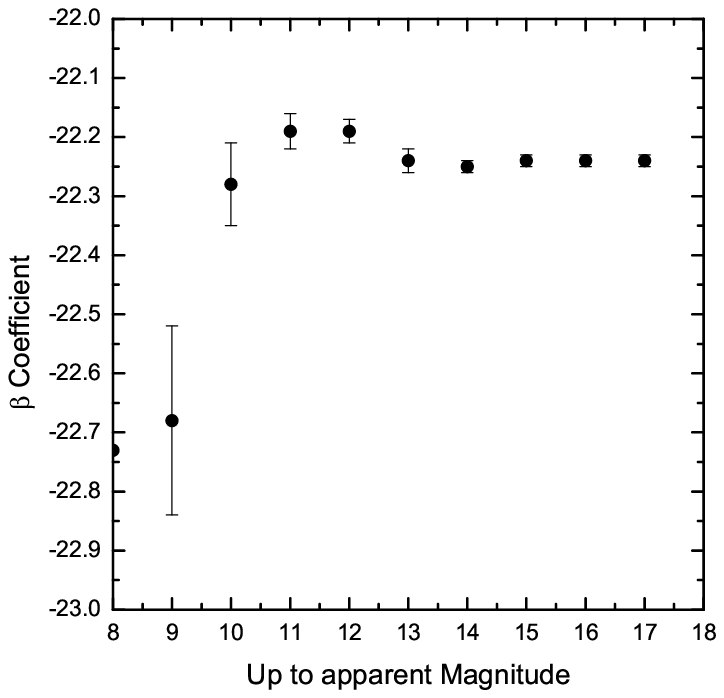}\caption{Variation
of the $\beta$ coefficient as a function of apparent magnitude.} \label{fig:beta2mag1}
\end{figure}

In Figures \ref{fig:alpha2mag1} and \ref{fig:beta2mag1} we present the variations of the coefficients $\alpha$ and $\beta$ as a function of
apparent magnitude up to the point to which the galaxies have been accumulated. It is interesting to note that the values of both coefficients
tend to a nearly constant value starting from apparent magnitude $i \sim +13$. This result suggests that gathering galaxies in fixed intervals of apparent magnitude may be the procedure that must
be followed in the calculation of the values of the coefficients of all the structural relations, not only for the Tully-Fisher relation, but also for the Kormendy relation, the Fundamental Plane and the Faber-Jackson relation, as long as the redshift interval covered is not too large, so that we need not be worried about galactic evolution effects.

\section{Conclusions}
\label{sec:conclusions}

In this preliminary work we have performed an investigation on the variation of the coefficients $(\alpha, \beta)$ of the TFR due to the way the galaxies used for this determination are grouped together. Again, we would like to stress the fact that the aim of this paper is not to obtain the true values of the coefficients of the TFR, but rather to show that their values may vary significantly due to observational restrictions. In order to do this, we have followed the directives presented in previous papers (Nigoche-Netro et al. 2008, 2009 \& 2010) and have grouped a sample of 2411 galaxies taken from the literature (Mathewson \& Ford, 1996) in several different ways which may be conceptualised as different observational groups of data. This sample of galaxies was chosen to perform this investigation because it consisted only of spiral galaxies which Mathewson \&  Ford had already identified as good candidates to be used in the TFR, and for which they presented all the necessary data required for our calculations; besides all the multiple corrections which are usually applied to samples of galaxies had already been applied, making this sample an ideal one for a straight forward use in our investigation.

In all the cases reported we found that the values of the coefficients vary more than the typical size of their corresponding error bars. A non-parametric statistical analysis of our results shows that we may reject the $null$ $hypothesis$, that states that the variations of the values of the coefficients is simply due to statistical fluctuations, with a high degree of confidence (see Table \ref{tab:hypothesis}) considering the size of the sample and the nature of our study.

In this preliminary, first-approach work, we have only shown that the values of the TFR may vary depending on how the sample of galaxies is collected. We do not, as yet, know whether these variations imply different physical properties between the alternate groups of galaxies used in the calculation of the TFR coefficients. However, we feel this not to be the case, because it appears that the differences in parameter values are due to three main causes: i) the way the different samples of galaxies are collected, here we certainly have the effects of observational cuts, ii) the form of the galaxy-distribution on the plane defined by the parameters of the relation in question (Tully-Fisher, Faber-Jackson, Fundamental Plane, and Kormendy). This point is fully explained in the Nigoche-Netro et al. (2008, 2009, 2010) papers and iii) the size of the intrinsic scatter of the relation. A detailed study with a larger sample of galaxies would shed more light into these points.

We also advanced the idea that the correct way of calculating the values of the coefficients of the structural relations for early type galaxies (KR, FPR, and FJR), as well as those for the TFR would be following the second iterative procedure presented in Section \ref{sec:magnitudecases} which accumulates galaxies in {\bf apparent magnitude} starting with the brightest ones and adding fainter ones in convenient steps of apparent magnitude.

\section*{Acknowledgments}

We thank the support provided by Instituto de Astronom\'{\i}a at Universidad Nacional Aut\'onoma de M\'exico (UNAM), and Instituto de Astronom\'{\i}a y Meteorolog\'{\i}a, Universidad de Guadalajara (UG). We would like to thank J. C. Yustis-Rubio for help with the figures.
We thank an anonymous referee for his/her comments, they have greatly improved the presentation of this paper. We also thank Direcci\'on General de Asuntos del Personal Acad\'emico, DGAPA at UNAM for financial support under projects number PAPIIT IN103813, IN111713 and IN102617. We acknowledge the usage of the HyperLeda database (http://leda.univ-lyon1.fr).

\vfill
\section{Appendix}
\label{sec:appendix}

In this appendix we shall present a detailed description of the different cases we studied.

\subsection{Case 1. Fixed Magnitude Intervals $(\Delta M=1)$ }
\label{sec:case1}

\begin{figure}
\begin{center}
\includegraphics[width=8.0cm,height=8.0cm]{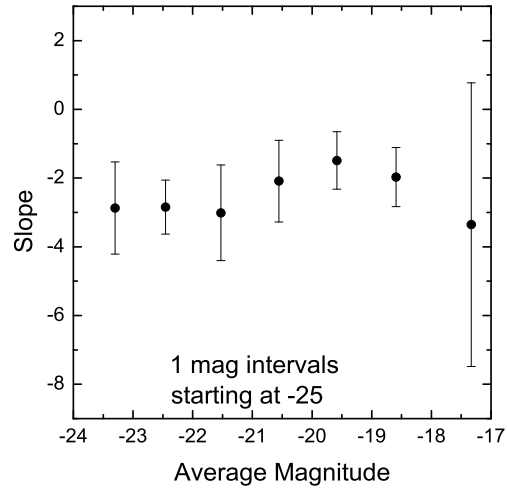}
\end{center}
\caption{Average values of the slope of the TFR for the fixed magnitude width $(\Delta M=1)$ case.} \label{fig:CASE1SLOPE}
\end{figure}

\begin{figure}
\begin{center}
\includegraphics[width=8.0cm,height=8.0cm]{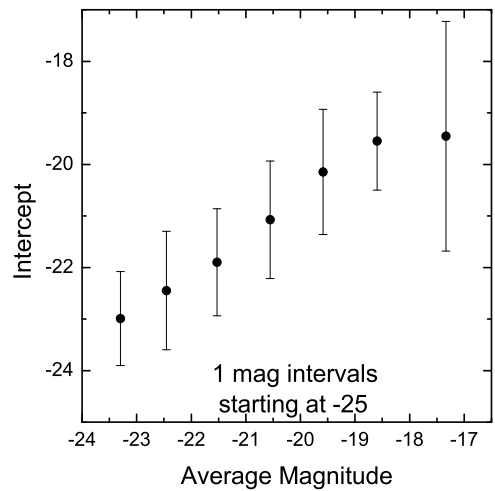}
\end{center}
\caption{Average values of the intercept of the TFR for the fixed magnitude width $(\Delta M=1)$ case.} \label{fig:CASE1INT}
\end{figure}

The magnitude range was divided into fixed magnitude intervals of 1 magnitude width (in bins of this width we ensure a sufficiently large number of galaxies in most cases), starting at $I=-25$ until $I=-13$. The average values of the slope $(\alpha)$ and the intercept $(\beta)$ of the TFR were calculated, and their values are presented in Figures~\ref{fig:CASE1SLOPE} and \ref{fig:CASE1INT}.

\subsection{Case 2. Increasing Width Magnitude Intervals (Increment=1 mag)}
\label{sec:case2}

\begin{figure}
\begin{center}
\includegraphics[width=8.0cm,height=8.0cm]{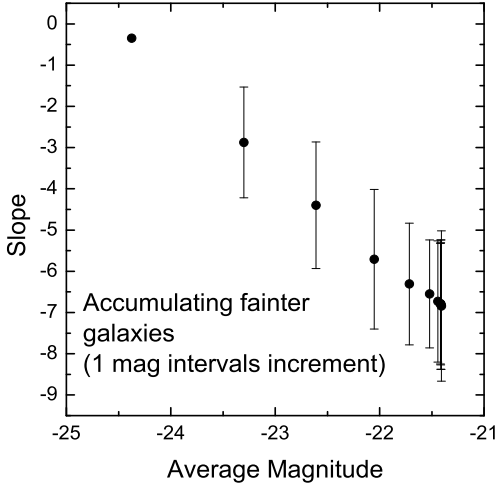}
\end{center}
\caption{Average values of the slope of the TFR for the increasing width magnitude interval $(\Delta M=1)$ case.} \label{fig:CASE2SLOPE}
\end{figure}

\begin{figure}
\begin{center}
\includegraphics[width=8.0cm,height=8.0cm]{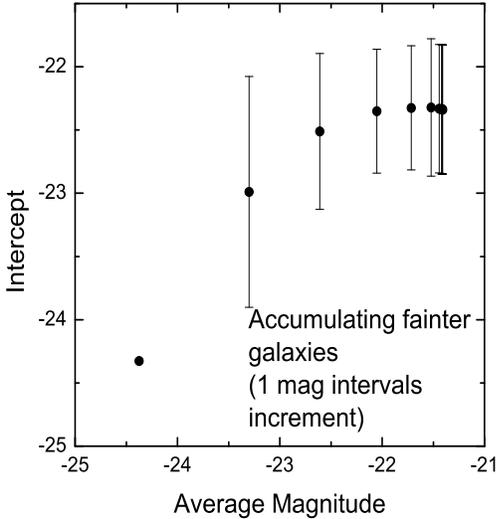}
\end{center}
\caption{Average values of the intercept of the TFR for the increasing width magnitude interval $(\Delta M=1)$ case.} \label{fig:CASE2INT}
\end{figure}

The magnitude range was divided into increasing width magnitude intervals starting at $I=-25$ until $I=-13$ and in each step we accumulated all the fainter galaxies in the next 1-mag interval, so we started with the interval $-25 \leq I \leq -24$, the next calculation was performed on the interval $-25 \leq I \leq  -23$ and the last interval was $-25 \leq I \leq -13$. The average values of the slope $(\alpha)$ and the intercept$(\beta)$ of the TFR calculated in this fashion are presented in Figures~\ref{fig:CASE2SLOPE} and \ref{fig:CASE2INT}.

\subsection{Case 3. Increasing Width Magnitude Intervals (Increment=1 mag)}
\label{sec:case3}

\begin{figure}
\begin{center}
\includegraphics[width=8.0cm,height=8.0cm]{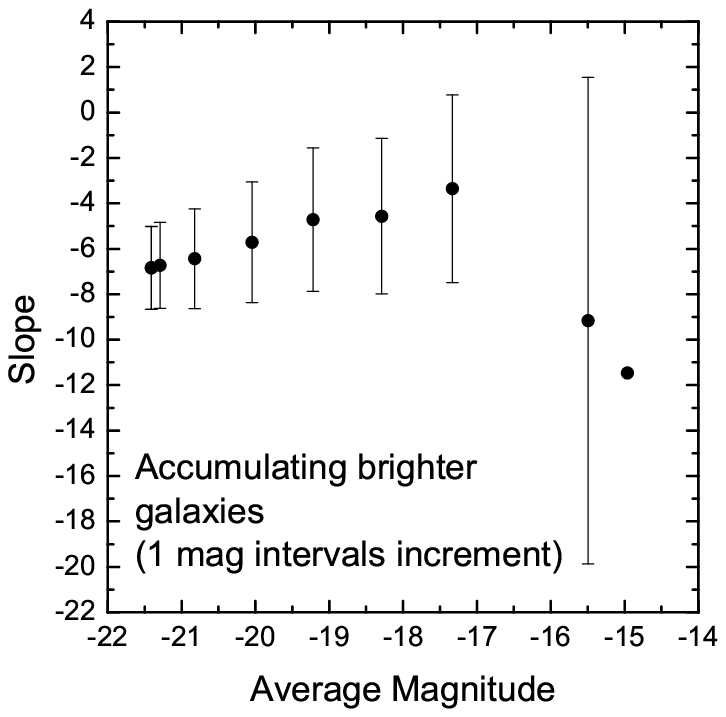}
\end{center}
\caption{Average values of the slope of the TFR for the increasing width magnitude interval $(\Delta M=-1)$ case.} \label{fig:CASE3SLOPE}
\end{figure}

\begin{figure}
\begin{center}
\includegraphics[width=8.0cm,height=8.0cm]{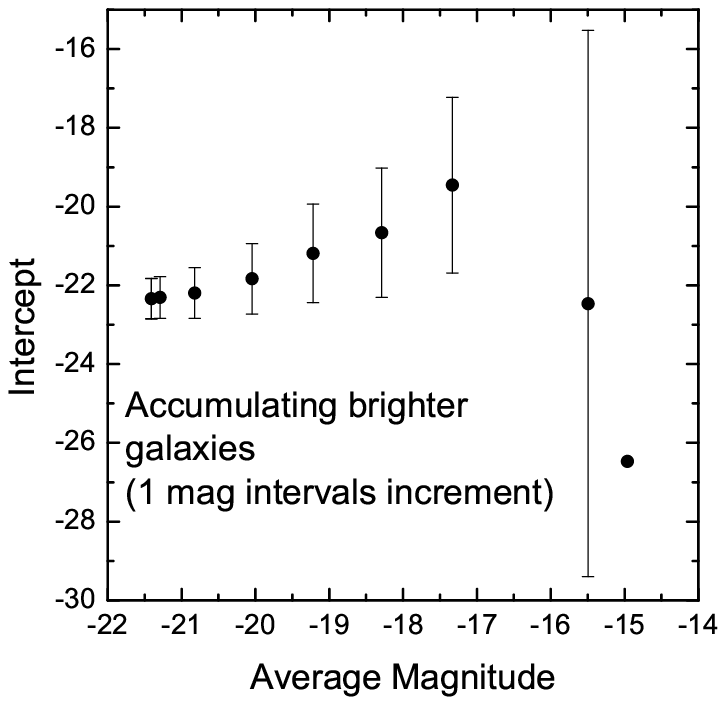}
\end{center}
\caption{Average values of the intercept of the TFR for the increasing width magnitude interval $(\Delta M=-1)$ case.} \label{fig:CASE3INT}
\end{figure}

The magnitude range was divided into increasing width magnitude intervals starting at $I=-13$ until $I=-25$ and in each step we accumulated all the brighter galaxies in the next 1-mag interval, so we started with the interval $-13 \geq I \geq -14$, the next calculation was performed on the interval $-13 \geq I \geq -15$ and the last interval was $-13 \geq I \geq -25$. The average values of the slope $(\alpha)$ and the intercept$(\beta)$ of the TFR calculated using this technique are shown in Figures~\ref{fig:CASE3SLOPE} and \ref{fig:CASE3INT}.

\subsection{Case 4. Increasing Width Magnitude Intervals (Increment=1 mag) starting at $I=-25.5$}
\label{sec:case4}

\begin{figure}
\begin{center}
\includegraphics[width=8.0cm,height=8.0cm]{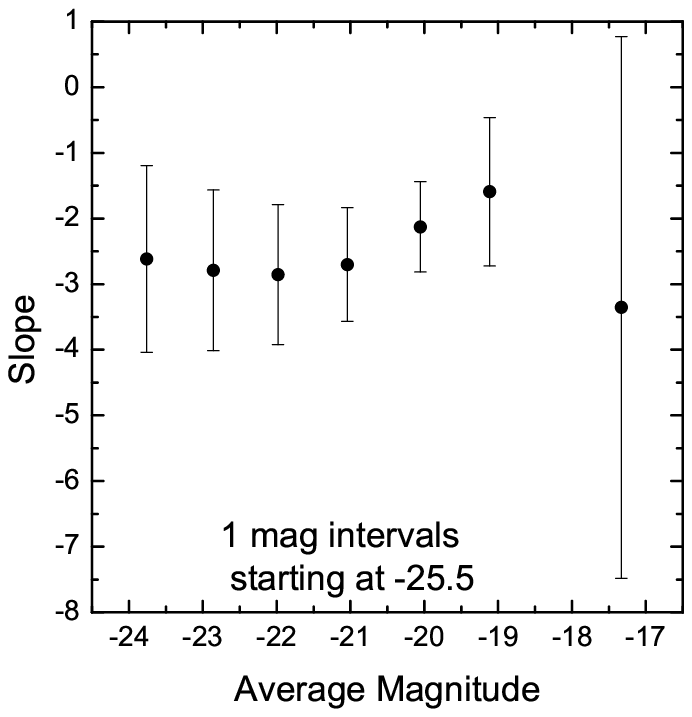}
\end{center}
\caption{Average values of the slope of the TFR for the fixed magnitude width $(\Delta M=1)$ case. Starting at $I=-10.5$ which is a different value from that in case 1.} \label{fig:CASE4SLOPE}
\end{figure}

\begin{figure}
\begin{center}
\includegraphics[width=8.0cm,height=8.0cm]{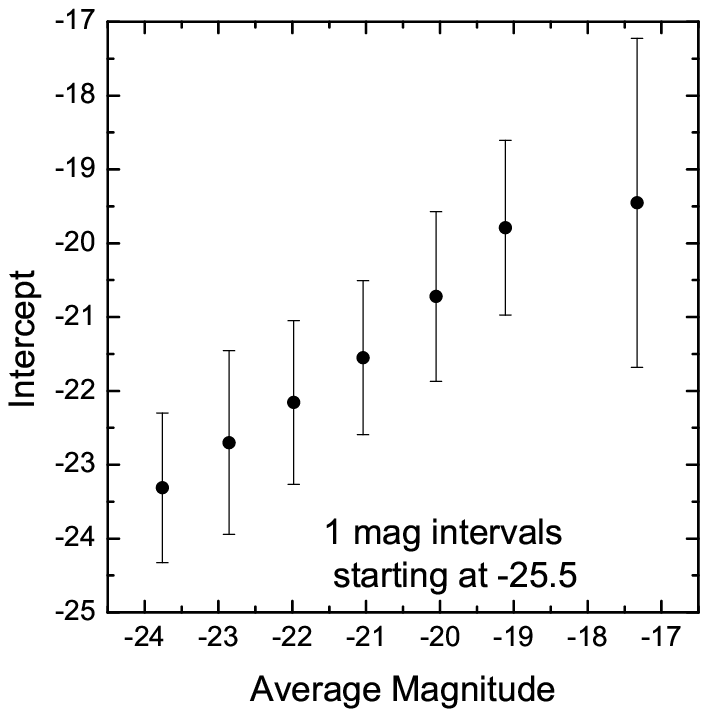}
\end{center}
\caption{Average values of the intercept of the TFR for the fixed magnitude width $(\Delta M=1)$ case. Starting at $I=-10.5$ which is a different value from that in case 1.} \label{fig:CASE4INT}
\end{figure}

In this case we divided the magnitude interval, as for case 1, in fixed magnitude intervals of width 1-mag. However, the limits of the intervals were different from those in case 1. In this case we started considering the galaxies from $I=-25.5$ and moved steadily in 1-magnitude intervals all the way to $I=-13.5$. The average values of the slope $(\alpha)$ and the intercept$(\beta)$ of the TFR calculated in this manner are presented in Figures~\ref{fig:CASE4SLOPE} and \ref{fig:CASE4INT}.

\subsection{Case 5. Fixed Magnitude Intervals $(\Delta M=2)$}
\label{sec:case5}

\begin{figure}
\begin{center}
\includegraphics[width=8.0cm,height=8.0cm]{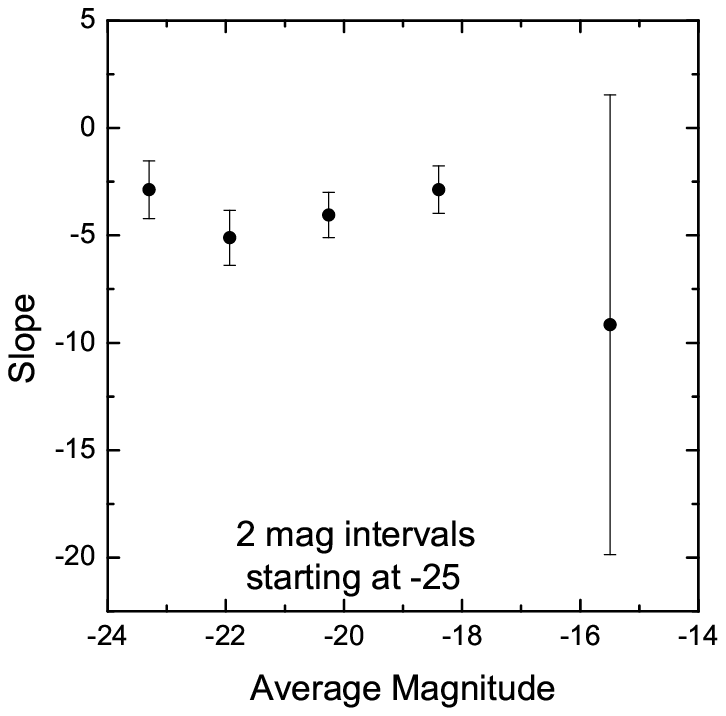}
\end{center}
\caption{Average values of the slope of the TFR for the increasing width magnitude interval $(\Delta M=2)$ case.} \label{fig:CASE5SLOPE}
\end{figure}

\begin{figure}
\begin{center}
\includegraphics[width=8.0cm,height=8.0cm]{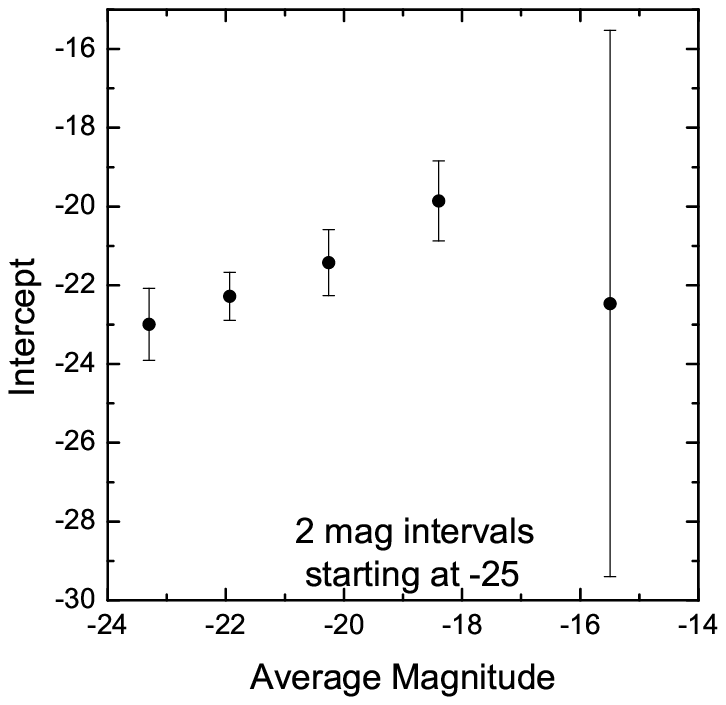}
\end{center}
\caption{Average values of the intercept of the TFR for the increasing width magnitude interval $(\Delta M=2)$ case.} \label{fig:CASE5INT}
\end{figure}

This case is rather similar to case 1, however the increment in magnitude is $\Delta M=2\ mag$ starting at $I=-25$ and finishing at $I=-13$. The first calculation is performed in the magnitude interval $-25 \leq I \leq -23$, whereas the last one is performed on the interval $-15 \leq I \leq -13$. The average values of the slope $(\alpha)$ and the intercept$(\beta)$ of the TFR calculated using this technique are shown in Figures~\ref{fig:CASE5SLOPE} and \ref{fig:CASE5INT}.

\bigskip

\begin{figure}
\begin{center}
\includegraphics[width=8.0cm,height=8.0cm]{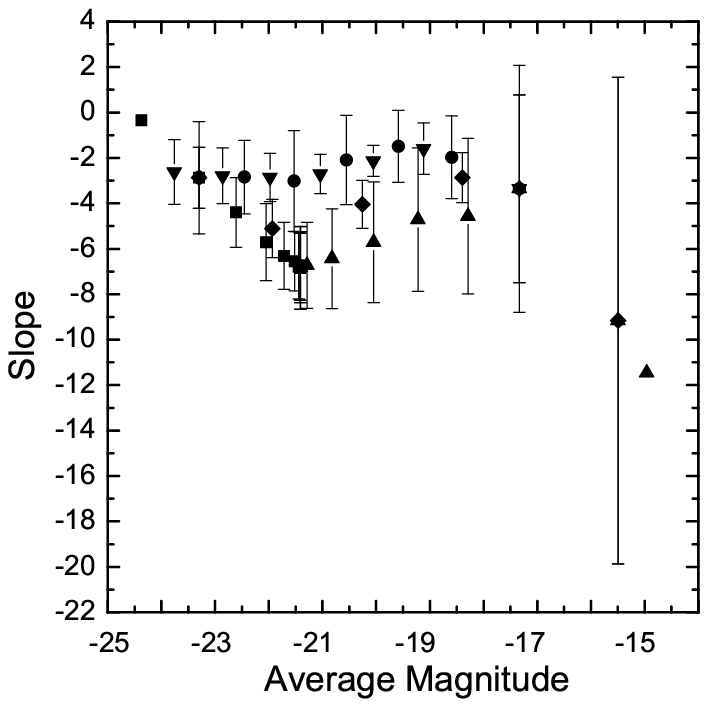}
\end{center}
\caption{Average values of the slope for the 5 cases discussed in section \ref{sec:cases}. Case 1: dots, Case 2: squares, Case 3: Triangles, Case 4: Inverted triangles and Case 5: Diamonds.} \label{fig:ALLSLOPE}
\end{figure}

\begin{figure}
\begin{center}
\includegraphics[width=8.0cm,height=8.0cm]{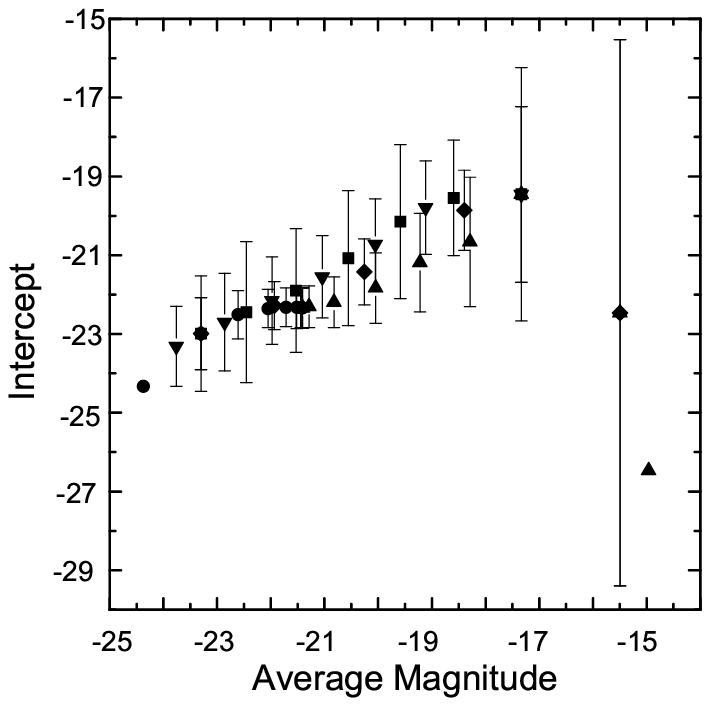}
\end{center}
\caption{Average values of the intercept for the 5 cases discussed in section \ref{sec:cases}. Case 1: dots, Case 2: squares, Case 3: Triangles, Case 4: Inverted triangles and Case 5: Diamonds.} \label{fig:ALLINTERCEPT}
\end{figure}

In the following two Figures \ref{fig:ALLSLOPE} and \ref{fig:ALLINTERCEPT}, we present, on the same plane, the values of the slope as well as those of the intercept for the five cases discussed above. These Figures allow us to appreciate visually the variation in the values of these parameters for the different cases.

In this appendix we present tables with information relevant to the calculation of the values of the parameters of the TFR. Each table corresponds to one of the cases discussed in Section \ref{sec:cases}.

\begin{table*}
\caption{$\alpha$ and $\beta$ values for Case 1 } \label{tab:tableCase1} \centering
\setlength{\tabcolsep}{0.5\tabcolsep}
\begin{tabular}{|c|c|c|c|}
\hline
Magnitude Interval          &    Number of Galaxies     &     Slope ($\alpha$)     &    Intercept ($\beta$) \\
\hline
                            &                           &                          &                        \\
$ -25 \leq M_I \leq -23 $   &     $144$                 &     $-2.8 \pm 1.3$       &     $-22.9 \pm 0.9$      \\
$ -23 \leq M_I \leq -22 $   &     $650$                 &     $-2.8 \pm 0.7$       &     $-22.4 \pm 1.1$      \\
$ -22 \leq M_I \leq -21 $   &     $845$                 &     $-3.0 \pm 1.3$       &     $-21.8 \pm 1.0$      \\
$ -21 \leq M_I \leq -20 $   &     $477$                 &     $-2.0 \pm 1.1$       &     $-21.0 \pm 1.1$      \\
$ -20 \leq M_I \leq -19 $   &     $212$                 &     $-1.4 \pm 0.8$       &     $-20.1 \pm 1.2$      \\
                            &                           &                          &                         \\
$ -19 \leq M_I \leq -18 $   &     $ 63$                 &     $-1.9 \pm 0.8$       &     $-19.5 \pm 0.9$      \\
$ -18 \leq M_I \leq -17 $   &     $ 20$                 &     $-3.3 \pm 4.1$       &     $-19.4 \pm 2.2$      \\
\hline
\end{tabular}
\end{table*}

\begin{table*}
\caption{$\alpha$ and $\beta$ values for Case 2 } \label{tab:tableCase2} \centering
\setlength{\tabcolsep}{0.5\tabcolsep}
\begin{tabular}{|c|c|c|c|}
\hline
Magnitude Interval          &    Number of Galaxies      &     Slope ($\alpha$)     &    Intercept ($\beta$) \\
\hline
                             &                            &                          &                        \\
$-25 \leq M_I \leq -24 $     &     $  2$                  &     $-0.3 \pm 0.0$       &     $-24.3 \pm 0.0$      \\
$-25 \leq M_I \leq -23 $     &     $144$                  &     $-2.8 \pm 1.3$       &     $-22.9 \pm 0.9$      \\
$-25 \leq M_I \leq -22 $     &     $794$                  &     $-4.3 \pm 1.5$       &     $-22.5 \pm 0.6$      \\
$-25 \leq M_I \leq -21 $     &     $1639$                 &     $-5.7 \pm 1.6$       &     $-22.3 \pm 0.4$      \\
$-25 \leq M_I \leq -20 $     &     $2116$                 &     $-6.3 \pm 1.4$       &     $-22.3 \pm 0.4$      \\
                             &                            &                          &                         \\
$-25 \leq M_I \leq -19 $     &     $2328$                 &     $-6.5 \pm 1.3$       &     $-22.3 \pm 0.5$      \\
$-25 \leq M_I \leq -18 $     &     $2391$                 &     $-6.7 \pm 1.4$       &     $-22.3 \pm 0.5$      \\
$-25 \leq M_I \leq -17 $     &     $2408$                 &     $-6.7 \pm 1.4$       &     $-22.3 \pm 0.5$      \\
$-25 \leq M_I \leq -16 $     &     $2409$                 &     $-6.7 \pm 1.4$       &     $-22.3 \pm 0.5$      \\
$-25 \leq M_I \leq -15 $     &     $2410$                 &     $-6.8 \pm 1.5$       &     $-22.3 \pm 0.5$      \\
                             &                            &                          &                        \\
$-25 \leq M_I \leq -14 $     &     $2410$                 &     $-6.8 \pm 1.5$       &     $-22.3 \pm 0.5$      \\
$-25 \leq M_I \leq -13 $     &     $2411$                 &     $-6.8 \pm 1.8$       &     $-22.3 \pm 0.5$      \\

\hline
\end{tabular}
\end{table*}

\begin{table*}
\caption{$\alpha$ and $\beta$ values for Case 3 } \label{tab:tableCase3} \centering
\setlength{\tabcolsep}{0.5\tabcolsep}
\begin{tabular}{|c|c|c|c|}
\hline
Magnitude Interval           &    Number of Galaxies      &     Slope ($\alpha$)     &    Intercept ($\beta$) \\
\hline
                               &                            &                          &                        \\
$ -13 \geq M_I \geq -14 $      &     $  1$                  &           $-$            &          $-$           \\
$ -13 \geq M_I \geq -15 $      &     $  1$                  &           $-$            &          $-$           \\
$ -13 \geq M_I \geq -16 $      &     $  2$                  &    $-11.4 \pm 0.0$       &    $-26.5 \pm 0.0$      \\
$ -13 \geq M_I \geq -17 $      &     $  3$                  &     $-9.1 \pm 10.7$      &     $-22.4 \pm 6.9$      \\
$ -13 \geq M_I \geq -18 $      &     $ 20$                  &     $-3.3 \pm 4.1$       &     $-19.4 \pm 2.2$      \\
                               &                            &                          &                         \\
$ -13 \geq M_I \geq -19 $      &     $ 83$                  &     $-4.5 \pm 3.4$       &     $-20.6 \pm 1.6$      \\
$ -13 \geq M_I \geq -20 $      &     $285$                  &     $-4.7 \pm 3.1$       &     $-21.1 \pm 1.2$      \\
$ -13 \geq M_I \geq -21 $      &     $772$                  &     $-5.7 \pm 2.6$       &     $-21.8 \pm 0.8$      \\
$ -13 \geq M_I \geq -22 $      &     $1617$                 &     $-6.4 \pm 2.1$       &     $-22.1 \pm 0.6$      \\
$ -13 \geq M_I \geq -23 $      &     $2267$                 &     $-6.7 \pm 1.8$       &     $-22.3 \pm 0.5$      \\
                               &                            &                          &                         \\
$ -13 \geq M_I \geq -24 $      &     $2409$                 &     $-6.8 \pm 1.8$       &     $-22.3 \pm 0.5$      \\
$ -13 \geq M_I \geq -25 $      &     $2411$                 &     $-6.8 \pm 1.8$       &     $-22.3 \pm 0.5$      \\

\hline
\end{tabular}
\end{table*}

\begin{table*}
\caption{$\alpha$ and $\beta$ values for Case 4 } \label{tab:tableCase4} \centering
\setlength{\tabcolsep}{0.5\tabcolsep}
\begin{tabular}{|c|c|c|c|}
\hline
Magnitude Interval              &    Number of Galaxies     &     Slope ($\alpha$)     &    Intercept ($\beta$) \\
\hline
                                  &                           &                          &                        \\
$ -24.5 \leq M_I \leq -23.5 $     &     $ 27$                 &     $-2.6 \pm 1.4$       &     $-23.3 \pm 1.0$      \\
$ -23.5 \leq M_I \leq -22.5 $     &     $400$                 &     $-2.7 \pm 1.2$       &     $-22.7 \pm 1.2$      \\
$ -22.5 \leq M_I \leq -21.5 $     &     $826$                 &     $-2.8 \pm 1.0$       &     $-22.1 \pm 1.1$      \\
$ -21.5 \leq M_I \leq -20.5 $     &     $670$                 &     $-2.7 \pm 0.8$       &     $-21.5 \pm 1.0$      \\
$ -20.5 \leq M_I \leq -19.5 $     &     $327$                 &     $-2.1 \pm 0.6$       &     $-20.7 \pm 1.1$      \\
                                  &                           &                          &                         \\
$ -19.5 \leq M_I \leq -18.5 $     &     $120$                 &     $-1.5 \pm 1.1$       &     $-19.7 \pm 1.1$      \\
$ -18.5 \leq M_I \leq -17.5 $     &     $ 41$                 &     $-4.0 \pm 4.1$       &     $-20.1 \pm 2.1$      \\
\hline
\end{tabular}
\end{table*}

\begin{table*}
\caption{$\alpha$ and $\beta$ values for Case 5 } \label{tab:tableCase5} \centering
\setlength{\tabcolsep}{0.5\tabcolsep}
\begin{tabular}{|c|c|c|c|}
\hline
Magnitude Interval          &    Number of Galaxies     &     Slope ($\alpha$)     &    Intercept ($\beta$) \\
\hline
                              &                           &                          &                        \\
$ -25 \leq M_I \leq -23 $     &     $144$                 &     $-2.8 \pm 1.3$       &     $-22.9 \pm 0.9$      \\
$ -23 \leq M_I \leq -21 $     &     $1495$                &     $-5.1 \pm 1.2$       &     $-22.2 \pm 0.6$      \\
$ -21 \leq M_I \leq -19 $     &     $689$                 &     $-4.0 \pm 1.0$       &     $-21.4 \pm 0.8$      \\
$ -19 \leq M_I \leq -17 $     &     $ 80$                 &     $-2.8 \pm 1.1$       &     $-19.8 \pm 1.0$      \\
$ -17 \leq M_I \leq -13 $     &     $  3$                 &     $-9.1 \pm 10.7$      &     $-22.4 \pm 6.9$      \\

\hline
\end{tabular}
\end{table*}

\begin{figure}
\centering
\includegraphics[width=8.0cm,height=8.0cm]{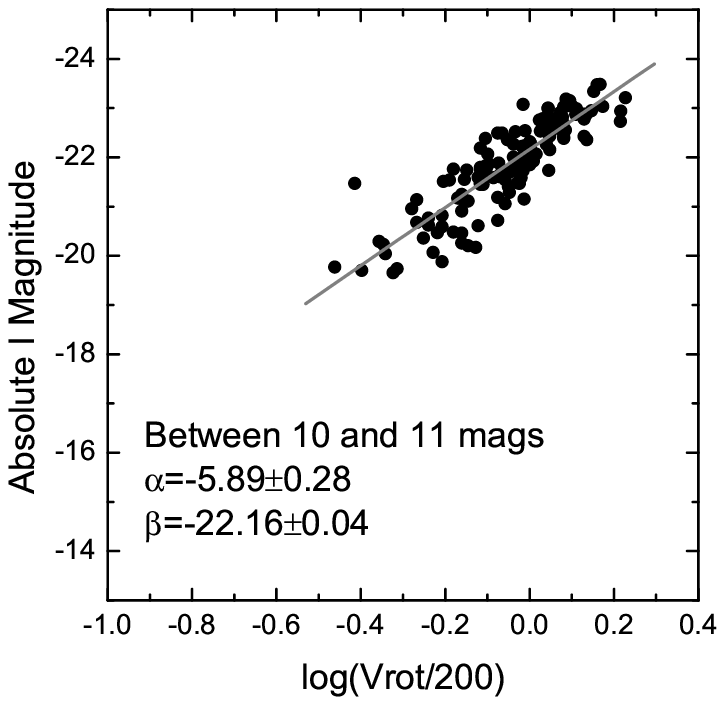}\caption{Galaxies
 with apparent magnitude between 10 and 11. Number of Galaxies=144.} \label{fig:between10and11}
\end{figure}

\begin{figure}
\centering
\includegraphics[width=8.0cm,height=8.0cm]{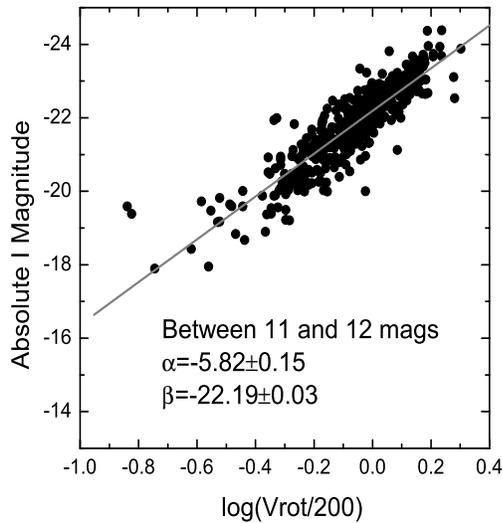}\caption{Galaxies
 with apparent magnitude between 11 and 12. Number of Galaxies=412.} \label{fig:between11and12}
\end{figure}

\begin{figure}
\centering
\includegraphics[width=8.0cm,height=8.0cm]{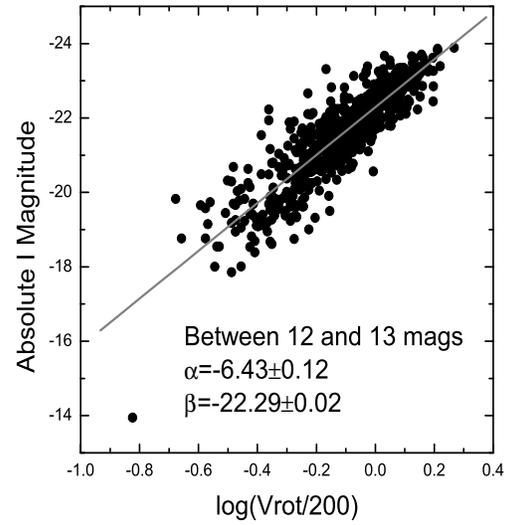}\caption{Galaxies
 with apparent magnitude between 12 and 13. Number of Galaxies=748.} \label{fig:between12and13}
\end{figure}

\begin{figure}
\centering
\includegraphics[width=8.0cm,height=8.0cm]{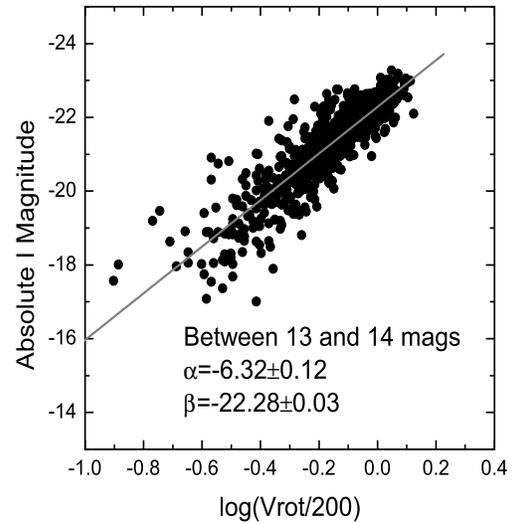}\caption{Galaxies
 with apparent magnitude between 13 and 14. Number of Galaxies=765.} \label{fig:between13and14}
\end{figure}

\begin{figure}
\centering
\includegraphics[width=8.0cm,height=8.0cm]{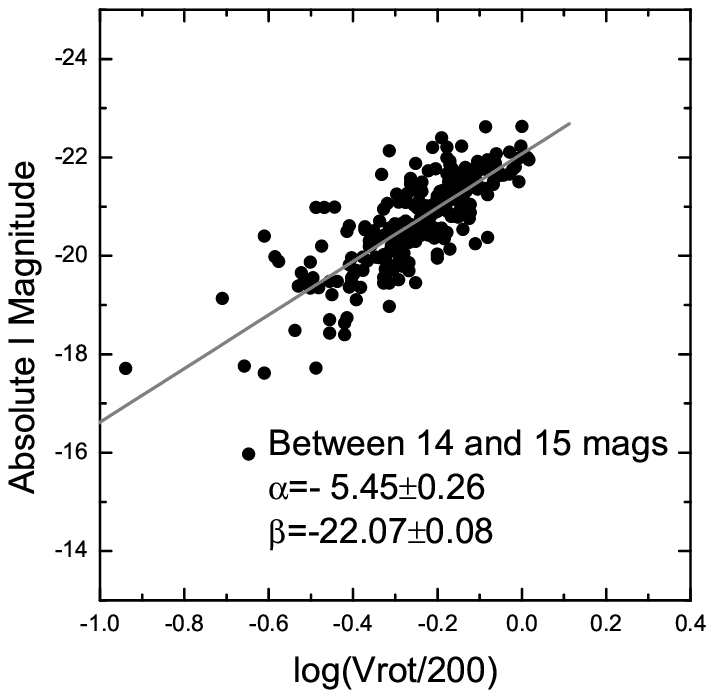}\caption{Galaxies
 with apparent magnitude between 14 and 15. Number of Galaxies=256.} \label{fig:between14and15}
\end{figure}

Now we shall present the corresponding graphs for Section \ref{sec:magnitudecases} (Figures \ref{fig:between10and11} to \ref{fig:between14and15}) of Absolute Magnitude vs $\log(V_{ROT}/200)$ for galaxies in our sample which are contained in specific intervals of
apparent magnitude. The figures corresponding to those intervals for which the number of galaxies is smaller than 50 are not presented in this paper because the least squares fit performed in these cases lacks enough statistical significance.

\begin{figure}
\centering
\includegraphics[width=8.0cm,height=8.0cm]{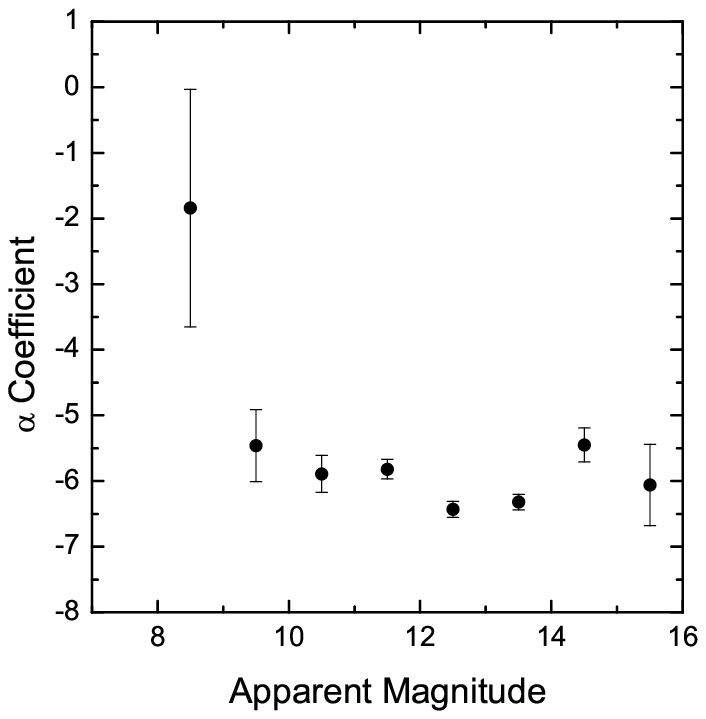}\caption{Variations of the value of the $\alpha$
coefficient for groups of galaxies contained in intervals 1 apparent magnitude wide.} \label{fig:alpha1mag}
\end{figure}

\begin{figure}
\centering
\includegraphics[width=8.0cm,height=8.0cm]{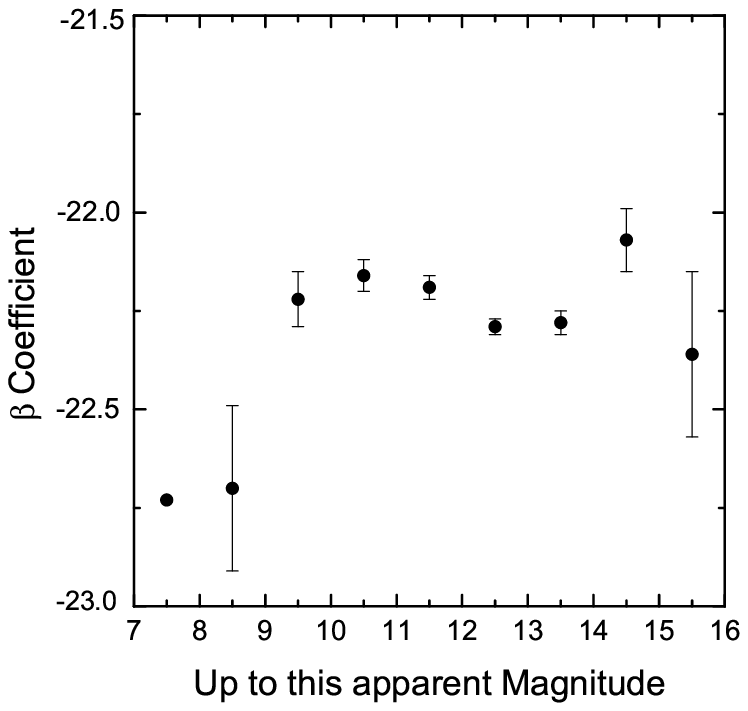}\caption{Variations of the value of the $\beta$
coefficient for groups of galaxies contained in intervals 1 apparent magnitude wide.} \label{fig:beta1mag}
\end{figure}

Now we present graphs of the variation of the $\alpha$ and $\beta$ coefficients as a function of the middle
point of the apparent magnitude intervals we chose to undertake this analysis. In Figure \ref{fig:alpha1mag} we see that the value
of $\alpha$ presents variations which are larger than the size of the error bars, however, its value stays contained within a
band that extends from $-5$ to $-7$. Figure \ref{fig:beta1mag} shows the values of the $\beta$ coefficient. This coefficient also
presents variations larger than the size of the typical error bars, but its value remains contained within a band extending from
$-22.0$ to $-22.5$.

\bigskip
At this time, we shall present a similar analysis in which galaxies will be accumulating, starting with those galaxies contained within a
bright apparent magnitude interval, until we reach the galaxies in the faintest apparent magnitude interval. We begin with the brightest galaxies in our sample up to an apparent magnitude of $+8$, continuing with all those galaxies with an apparent magnitude fainter than $+9$, until we reach
the total sample at an apparent magnitude of $+17$.

\begin{figure}
\centering
\includegraphics[width=8.0cm,height=8.0cm]{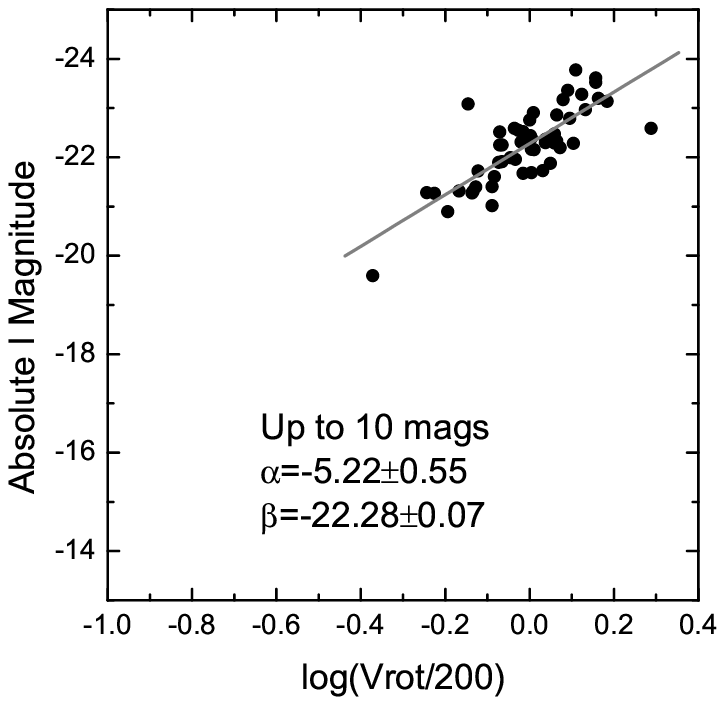}\caption{Galaxies
contained up to an apparent magnitude of $+10$. Number of Galaxies=54.} \label{fig:HASTA10}
\end{figure}

\begin{figure}
\centering
\includegraphics[width=8.0cm,height=8.0cm]{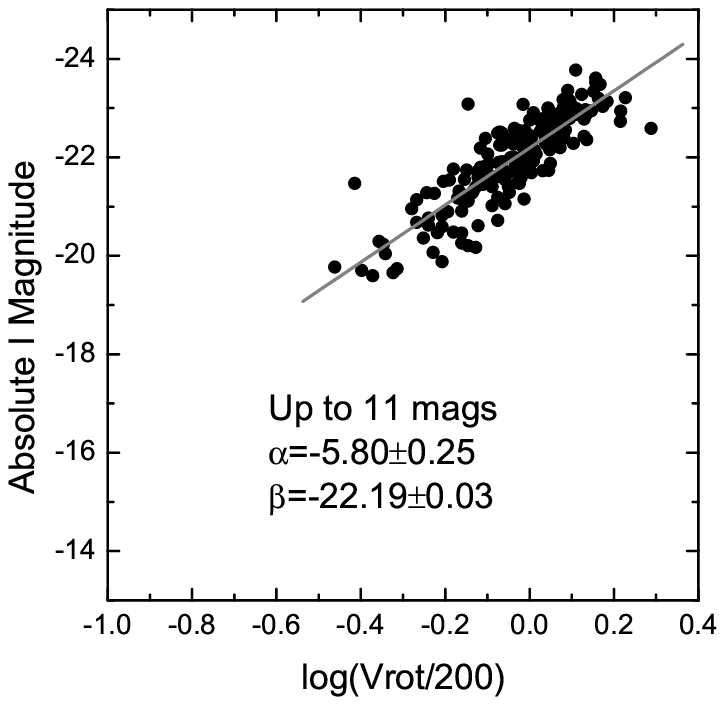}\caption{Galaxies
contained up to an apparent magnitude of $+11$. Number of Galaxies=198.} \label{fig:HASTA11}
\end{figure}

\begin{figure}
\centering
\includegraphics[width=8.0cm,height=8.0cm]{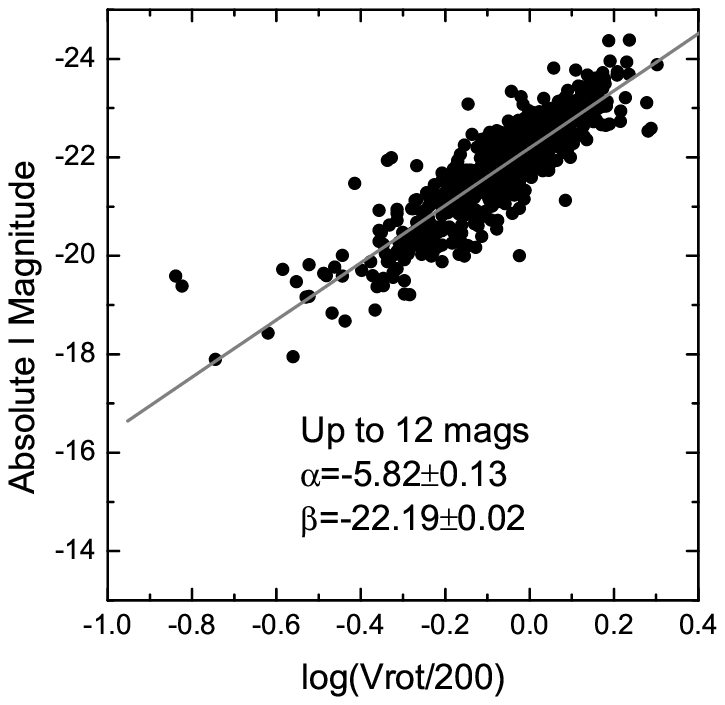}\caption{Galaxies
contained up to an apparent magnitude of $+12$. Number of Galaxies=610.} \label{fig:HASTA12}
\end{figure}

\begin{figure}
\centering
\includegraphics[width=8.0cm,height=8.0cm]{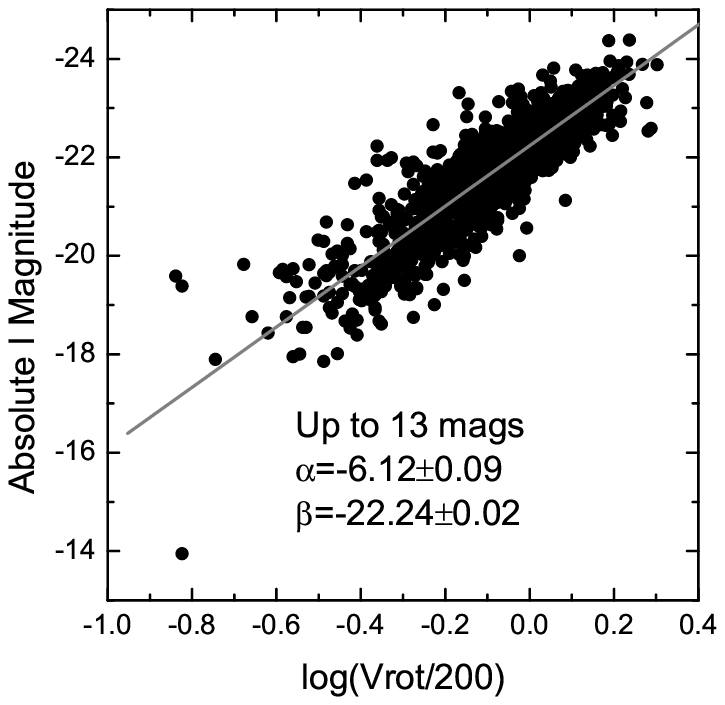}\caption{Galaxies
contained up to an apparent magnitude of $+13$. Number of Galaxies=1358.} \label{fig:HASTA13}
\end{figure}

\begin{figure}
\centering
\includegraphics[width=8.0cm,height=8.0cm]{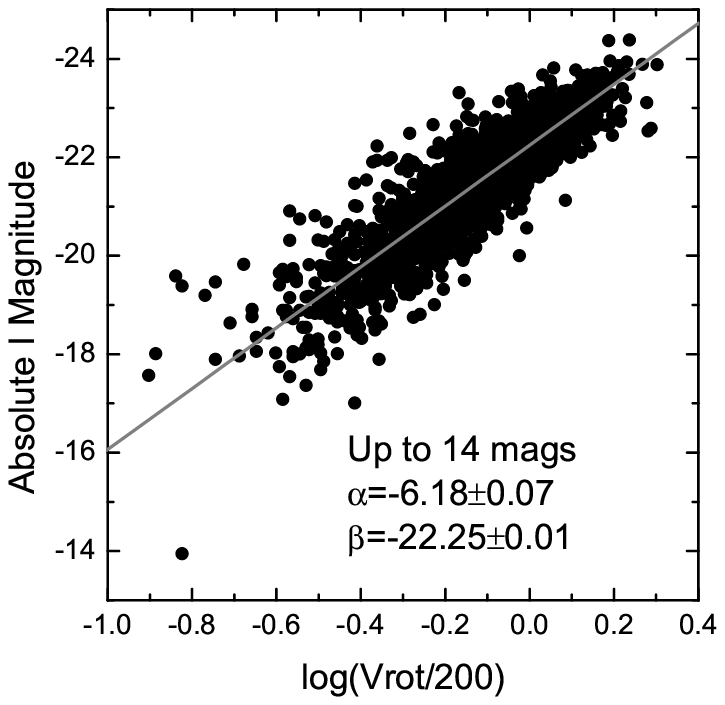}\caption{Galaxies
contained up to an apparent magnitude of $+14$. Number of Galaxies=2123.} \label{fig:HASTA14}
\end{figure}

\begin{figure}
\centering
\includegraphics[width=8.0cm,height=8.0cm]{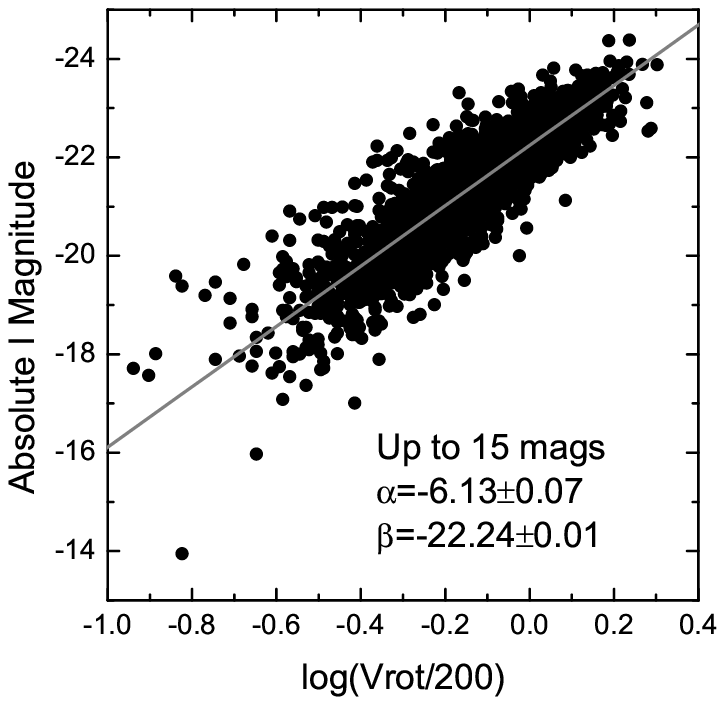}\caption{Galaxies
contained up to an apparent magnitude of $+15$. Number of Galaxies=2379.} \label{fig:HASTA15}
\end{figure}

\begin{figure}
\centering
\includegraphics[width=8.0cm,height=8.0cm]{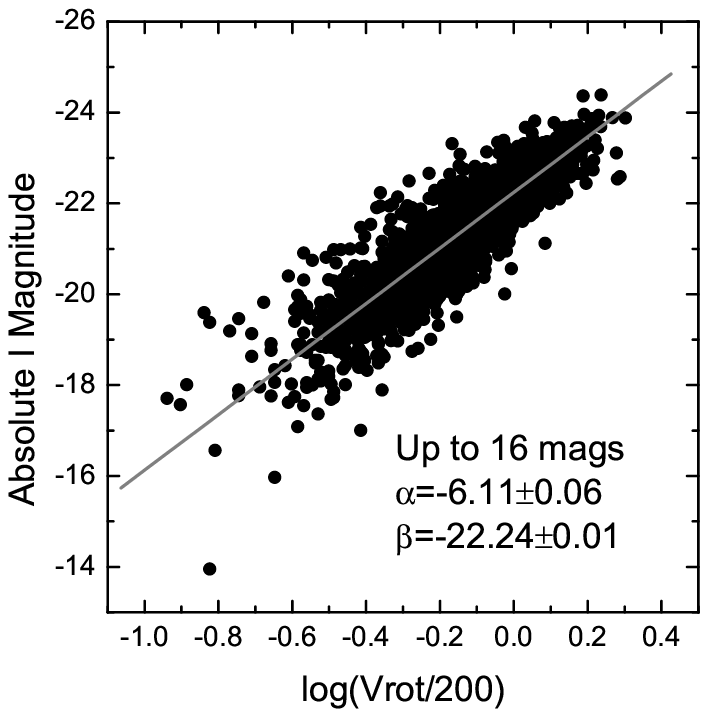}\caption{Galaxies
contained up to an apparent magnitude of $+16$. Number of Galaxies=2410.} \label{fig:HASTA16}
\end{figure}

\begin{figure}
\centering
\includegraphics[width=8.0cm,height=8.0cm]{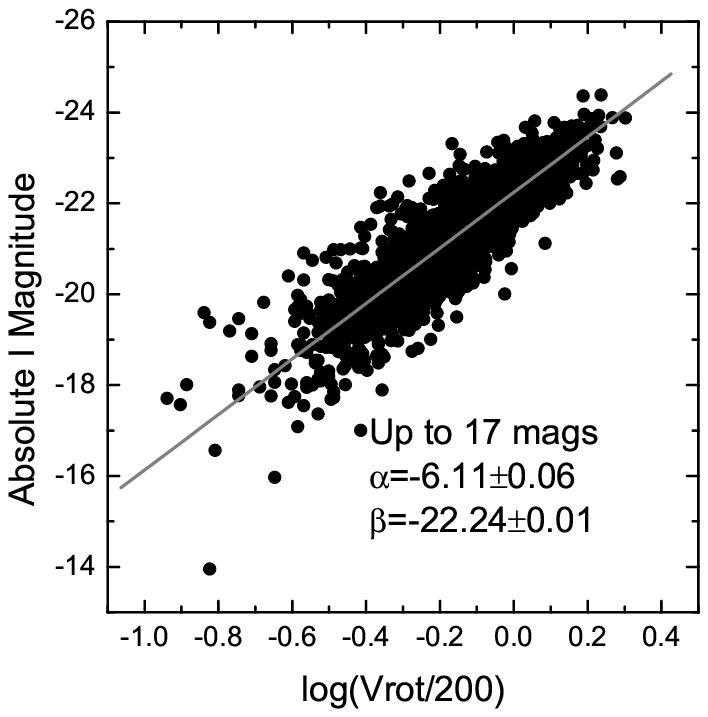}\caption{Galaxies
contained up to an apparent magnitude of $+17$. Number of Galaxies=2411.} \label{fig:HASTA17}
\end{figure}

In Figures \ref{fig:HASTA10} to \ref{fig:HASTA17} we present graphs of the fitting process to the galaxies in our sample as they accumulate
from bright to faint apparent magnitude intervals. Again, the figures corresponding to fits with a number of galaxies smaller than 50 are not presented.

\begin{figure}
\centering
\includegraphics[width=8.0cm,height=8.0cm]{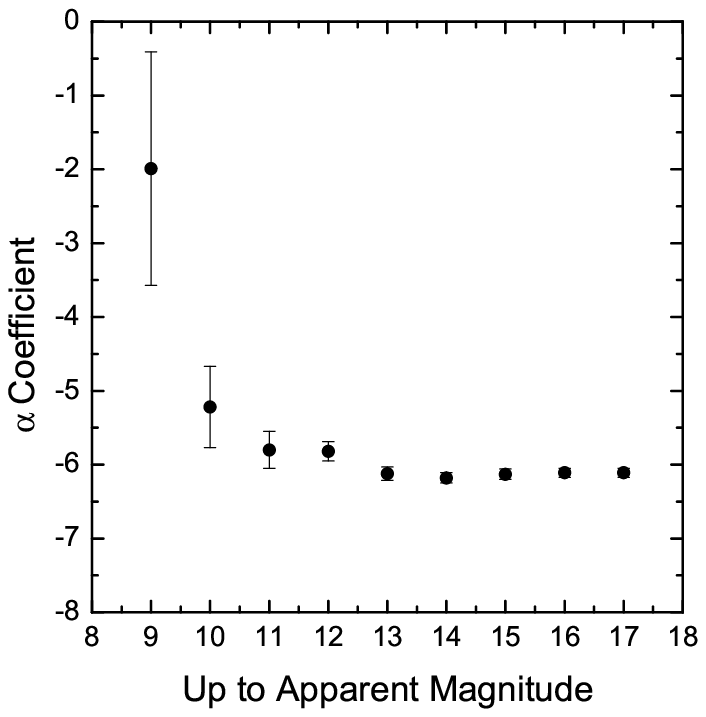}\caption{Variation of the $\alpha$ coefficient as a function of apparent magnitude.} \label{fig:alpha2mag}
\end{figure}

\begin{figure}
\centering
\includegraphics[width=8.0cm,height=8.0cm]{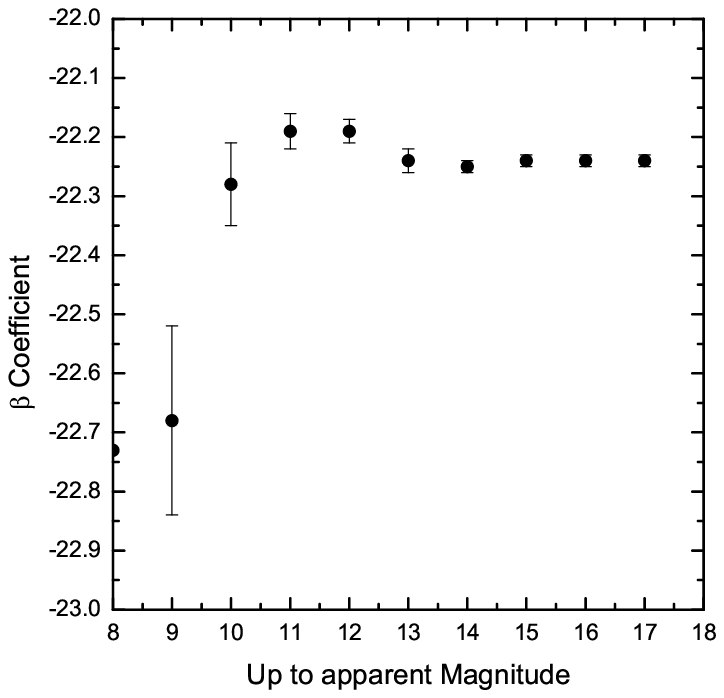}\caption{Variation
of the $\beta$ coefficient as a function of apparent magnitude.} \label{fig:beta2mag}
\end{figure}

In Figures \ref{fig:alpha2mag} and \ref{fig:beta2mag} we present the variations of the coefficients $\alpha$ and $\beta$ as a function of
apparent magnitude up to the point to which the galaxies have been accumulated. It is interesting to note that the values of both coefficients
tend to a nearly constant value starting from apparent magnitude $i \sim +13$. This result suggests that gathering galaxies in fixed intervals of apparent magnitude may be the procedure that must
be followed in the calculation of the values of the coefficients of all the structural relations, not only for the Tully-Fisher relation, but also for the Kormendy relation, the Fundamental Plane and the Faber-Jackson relation, as long as the redshift interval covered is not too large, so that we need not be worried about galactic evolution effects.

\end{document}